%% file: SingleN_EFT.tex
\documentclass[11pt,a4paper]{article}
\usepackage{jheppub}
\pdfoutput=1

\usepackage[utf8]{inputenc}
\usepackage{amssymb}
\usepackage{amsmath}
\usepackage{amsfonts}
\usepackage{graphicx}
\usepackage{xcolor}
\usepackage{xspace}
\usepackage[normalem]{ulem} 
\usepackage{lscape}
\usepackage{ wasysym }
\usepackage{float}
\usepackage{mathrsfs}
\usepackage{slashed}
\usepackage{multirow,rotating}

\usepackage{hyperref}
\hypersetup{
  colorlinks=true,
  allcolors=darkpurple,
  pdfborder={0 0 0},
  linktocpage=true
}

\definecolor{darkpurple}{rgb}{0.5,0,0.5}
\definecolor{cambridgeblue}{rgb}{0.64, 0.76, 0.68}
\definecolor{darkraspberry}{rgb}{0.53, 0.15, 0.34}

\def\gsim{\raise0.3ex\hbox{$\;>$\kern-0.75em\raise-1.1ex\hbox{$\sim\;$}}}
\def\lsim{\raise0.3ex\hbox{$\;<$\kern-0.75em\raise-1.1ex\hbox{$\sim\;$}}}

\newcommand{\ba}[1]{\begin{eqnarray} \label{(#1)}}
\newcommand{\ea}{\end{eqnarray}}

\def\gsim{\raise0.3ex\hbox{$\;>$\kern-0.75em\raise-1.1ex\hbox{$\sim\;$}}}
\def\lsim{\raise0.3ex\hbox{$\;<$\kern-0.75em\raise-1.1ex\hbox{$\sim\;$}}}

\makeatletter
\g@addto@macro\bfseries{\boldmath}
\makeatother

	\preprint{\begin{flushright} FTUV-21-1026.4809	\\IFIC/21-40 
	\end{flushright}}	

\title{Long-lived heavy neutral leptons at the LHC: four-fermion
          single-$N_R$ operators}

\author[a]{Rebeca Beltr\'an,}
\emailAdd{rebeca.beltran@ific.uv.es}
\affiliation[a]{{\it AHEP Group, Instituto de F\'{\i}sica Corpuscular --	CSIC/Universitat de Val{\`e}ncia, Apartado 22085,
	E--46071 Val{\`e}ncia, Spain}}

\author[b,c]{Giovanna Cottin,}
\emailAdd{giovanna.cottin@uai.cl}
\affiliation[b]{Departamento de Ciencias, Facultad de Artes Liberales, 
	Universidad Adolfo Ib\'a\~{n}ez,
	Diagonal Las Torres 2640, Santiago, Chile}
\affiliation[c]{Millennium Institute for Subatomic Physics at the High Energy Frontier (SAPHIR), Fernández Concha 700, Santiago, Chile}

\author[d,c]{Juan Carlos Helo,}
\emailAdd{jchelo@userena.cl}
\affiliation[d]{Departamento de F\' isica, Facultad de Ciencias, Universidad de La Serena, 
	Avenida Cisternas 1200, La Serena, Chile }

\author[a]{Martin Hirsch,}
\emailAdd{mahirsch@ific.uv.es}

\author[e]{Arsenii Titov,}
\emailAdd{arsenii.titov@ific.uv.es}
\affiliation[e]{{\it Departament de F{\'i}sica Te{\`o}rica, Universitat de Val{\`e}ncia and Instituto de F\'{\i}sica Corpuscular -- CSIC/Universitat de Val{\`e}ncia, Dr.~Moliner 50, E--46100 Burjassot, Spain}}

\author[f,g]{Zeren Simon Wang}
\emailAdd{wzs@mx.nthu.edu.tw}
\affiliation[f]{Department of Physics, National Tsing Hua University, Hsinchu 300, Taiwan}
\affiliation[g]{Center for Theory and Computation, National Tsing Hua University, Hsinchu 300, Taiwan}

\abstract{Interest in searches for heavy neutral leptons (HNLs) at the
  LHC has increased considerably in the past few years.  In the
  minimal scenario, HNLs are produced and decay via their mixing with
  active neutrinos in the Standard Model (SM) spectrum.  However, many
  SM extensions with HNLs have been discussed in the literature, which
  sometimes change expectations for LHC sensitivities drastically.  In
  the $N_R$SMEFT, one extends the SM effective field theory with
  operators including SM singlet fermions, which allows to study HNL
  phenomenology in a ``model independent'' way.  In this paper, we
  study the sensitivity of ATLAS to HNLs in the $N_R$SMEFT for
  four-fermion operators with a single HNL.  These operators might
  dominate both production and decay of HNLs, and we find that new
  physics scales in excess of 20~TeV could be probed at the
  high-luminosity LHC.}

\begin{document}
\maketitle

%


\input{tex/intro}

\input{tex/eft}

\input{tex/simulation}

\input{tex/results}

\input{tex/summary}

\bigskip
\section*{Acknowledgements}

\medskip

This work is supported by the Spanish grants PID2020-113775GB-I00
(AEI/10.13039/ 501100011033) and PROMETEO/2018/165 (Generalitat
Valenciana). R.B. acknowledges financial support from the Generalitat
Valenciana (grant ACIF/2021/052) and CSIC (JAE\\ICU-20-IFIC-2).
G.C. acknowledges support from ANID FONDECYT-Chile grant No. 3190051.
G.C. and J.C.H. also acknowledge support from grant ANID
FONDECYT-Chile No. 1201673 and ANID – Millennium Science Initiative
Program ICN2019\_044.  The work of A.T. is supported by the “Generalitat
Valenciana” under grant PROMETEO/2019/087, 
as well as by the FEDER/MCIyU-AEI grant FPA2017-84543-P 
and the AEI-MICINN grant PID2020-113334GB-I00 (AEI/10.13039/501100011033).
Z.S.W. is supported by the Ministry of Science
and Technology (MoST) of Taiwan with grant numbers
MoST-109-2811-M-007-509 and MoST-110-2811-M-007-542-MY3.

\bibliographystyle{JHEP}
\bibliography{RefsEFT}

\end{document}

%% file: tex/intro.tex

\section{Introduction}\label{sect:intro}

Interest in long-lived particles (LLPs) has grown largely in the last
few years~\cite{Alimena:2019zri,Lee:2018pag,Curtin:2018mvb}.  Many
models for LLPs have been discussed in the literature, most of which
are motivated by either dark matter or neutrino masses.  Heavy neutral
leptons (HNLs) are the prime example for LLPs connected with the
neutrino masses. HNLs are Standard Model (SM) singlet fermions that
couple to SM particles via their mixing with active neutrinos.

The minimal model that can realize this effective setup is the seesaw
mechanism, in which right-handed Majorana neutrinos, $N_R$, are added
to the SM particle content~\cite{Minkowski:1977sc,Yanagida:1979as,
  GellMann:1980vs,Mohapatra:1979ia,Schechter:1980gr}.  However, many
SM extensions, that aim to explain observed neutrino
data~\cite{deSalas:2020pgw} (see also Refs.
\cite{Capozzi:2021fjo,Esteban:2020cvm}) via electroweak scale
variants of the classical
seesaw~\cite{Mohapatra:1986bd,Bernabeu:1987gr,Akhmedov:1995ip,Akhmedov:1995vm},
do not include only HNLs.  For example, right-handed neutrinos appear
necessarily in left-right (LR) symmetric extension of the SM as the
neutral component of the right-lepton
doublet~\cite{Mohapatra:1974hk,Senjanovic:1975rk}.  If the additional
non-SM states, such as the $W_R$ and $Z'$ in the LR model, have masses
which are too large to be produced on-shell at the LHC, their effects
on HNL phenomenology is best treated in effective field theory (EFT).

The EFT of the SM, SMEFT, (see Ref.~\cite{Brivio:2017vri} for a
review) is a well-established framework in LHC searches (for global
analyses of collider data in this framework, see
Refs.~\cite{Ellis:2020unq,Ethier:2021bye}).  The extension of the
SMEFT to include right-handed neutrinos is called
$N_R$SMEFT.\footnote{In the literature, sometimes also called
  $\nu_R$SMEFT.}  This EFT has been originally discussed in Refs
\cite{delAguila:2008ir,Aparici:2009fh} and has attracted significant
interest in the last few years, from both
theoretical~\cite{Bhattacharya:2015vja,Liao:2016qyd,Li:2021tsq,Chala:2020vqp,Chala:2020pbn,Datta:2020ocb,Datta:2021akg}
and
phenomenological~\cite{Bischer:2019ttk,Alcaide:2019pnf,Butterworth:2019iff,Biekotter:2020tbd,Dekens:2020ttz,Han:2020pff,Li:2020lba,Li:2020wxi,DeVries:2020jbs,Cottin:2021lzz}
perspectives. Effective operators in the $N_R$SMEFT are now known up
to dimension $d=9$~\cite{Li:2021tsq}.  Phenomenological interest in
this EFT is motivated by the future upgrades of the LHC on one side
and the improvement in the sensitivities of low-energy experiments on
the other.

Effective interactions of $d \leq 6$ are the most interesting from a
phenomenological point of view. There are two $d = 5$ operators
involving $N_R$.  Their phenomenology has been studied in detail in
Refs.~\cite{Aparici:2009fh,Caputo:2017pit,Barducci:2020icf}.  The
$d=6$ operators containing $N_R$ can be divided into two classes: (i)
operators with two fermions and bosons and (ii) four-fermion
operators. The second class, in turn, can be partitioned into
operators with two $N_R$'s and operators with a single $N_R$.\footnote{There is also a lepton-number-violating operator with four
  $N_R$'s, but it requires at least two generations of HNLs. }  The LLP
phenomenology of pair operators has recently been studied in
Ref.~\cite{Cottin:2021lzz}.  Here, we will concentrate on operators
with a single $N_R$.  The phenomenology of single-$N_R$ operators is
decidedly different from that of pair operators.  First, pair
operators do not by themselves lead to decays of (the lightest) $N_R$.
Instead, for these operators $N_R$ decays are controlled by the mixing with active neutrinos.
This is different from the single-$N_R$ operators, which will usually
dominate the decay length of the HNLs in those parts of parameter
space where the operators are large enough to dominate $N_R$
production.  Thus, the parameter space that can be explored for these
two types of operators is very different, see Sec.~\ref{sect:res}.
Second, pair operators do not produce prompt charged leptons, except
in the parameter region where the decay length of the $N_R$ is so
short that the lepton from a $N_R$ decay is confused with a charged
lepton produced directly from $pp$ collisions at the interaction point (IP).  In all lepton-number-conserving single-$N_R$
operators, on the other hand, $N_R$'s are accompanied by a prompt
lepton (either a neutrino or a charged lepton).  This affects the
search strategy for the different operators.

Ref.~\cite{DeVries:2020jbs} studied single-$N_R$ operators for various proposed LLP ``far'' detectors, such as
MATHUSLA~\cite{Chou:2016lxi,Curtin:2018mvb,Alpigiani:2020tva},
CODEXb~\cite{Gligorov:2017nwh}, AL3X~\cite{Gligorov:2018vkc},
FASER~\cite{Feng:2017uoz}, and ANUBIS~\cite{Bauer:2019vqk}, as well as ATLAS, for HNLs produced from charm and bottom meson decays and hence with mass below 5 GeV.\footnote{For the expectations for these experiments in the minimal HNL scenario with only active-sterile neutrino mixing, see for example~Refs.~\cite{Helo:2018qej,Dercks:2018wum,Hirsch:2020klk}.}
In addition, Ref.~\cite{Zhou:2021ylt} very recently worked on phenomenology of the same set of single-$N_R$ operators associated with the third-generation leptons at Belle II, for HNLs produced from $\tau$ lepton decays.
For these reasons, in our numerical simulation we concentrate on ATLAS for heavier HNLs, and a short discussion will also be given for the expectations for CMS (see Sec.~\ref{sect:res}).

The rest of this paper is organized as follows.  In the next section,
we will discuss briefly $N_R$SMEFT at $d=6$.  This section
also entails a short discussion on how the single-$N_R$ operators
could be the low-energy remnant of some leptoquark or two Higgs
doublet models.  Sec.~\ref{sect:sim} discusses the details of the 
simulation we perform for the ATLAS detector.  In Sec.~\ref{sect:res},
we present our numerical results.  First, we discuss again briefly the
minimal case, in which HNLs are produced and decay via mixing only.
While this was previously done by some of us in
Ref.~\cite{Cottin:2018nms}, we now also simulate the expectations for
HNLs coupled to $\tau$'s, including both neutral and charged currents
leading to more realistic estimates for the future ATLAS
sensitivities.  We then present our results for the different
single-$N_R$ operators.  Cross sections and decay lengths depend on
both, operator type and generation indices in the SM sector.  For the
first generation of SM quarks, sensitivities will reach new physics
scales in excess of 20~TeV at the high-luminosity LHC.  We then close
with a short summary of our results.

%% file: tex/eft.tex

\section{Effective theory with $N_R$}\label{sect:eft}

\subsection{Effective interactions}

In this section, we briefly introduce the $N_R$SMEFT, focusing on the
operators of interest for the current work.  If HNLs with masses below
or around the electroweak scale exist in nature, the effects of new
multi-TeV physics at much smaller energies can be systematically
described in terms of an EFT built out of the SM fields and $N_R$.  At
renormalizable level, in addition to the SM operators, there are a
Majorana mass term for $N_R$ and a $d=4$ operator describing the
fermion portal:
\begin{equation}
 \mathcal{L}_\mathrm{ren} = \mathcal{L}_\mathrm{SM} + \overline{N_R} i \slashed{\partial} N_R 
 - \left[\frac{1}{2} \overline{N_R^c} M_N N_R + \overline{L} \tilde{H} Y_N N_R + \text{h.c.}\right],
\end{equation}
where $L$ stands for the SM lepton doublets, $H$ is the Higgs doublet
($\tilde{H} = \epsilon H^\ast$, $\epsilon$ is the totally
antisymmetric tensor), and $N_R^c \equiv C \overline{N_R}^T$ with $C$
being the Dirac charge conjugation matrix.  The Majorana mass matrix
$M_N$ is a symmetric $n_N \times n_N$ matrix, with $n_N$ denoting the
number of HNL generations, and $Y_N$ is a generic $3 \times n_N$
matrix of Yukawa couplings.

Upon including non-renormalizable interactions $\mathcal{O}_i^{(d)}$
with $d \geq 5$, the full Lagrangian reads
\begin{equation}
 \mathcal{L} = \mathcal{L}_\mathrm{ren} + \sum_{d \geq 5} \frac{1}{\Lambda^{d-4}} \sum_i c_i^{(d)} \mathcal{O}_i^{(d)}\,,
\end{equation}
where $c_i^{(d)}$ are the Wilson coefficients, and the second sum goes
over all independent interactions at a given dimension $d$.  At $d=5$,
in addition to the renowned Weinberg operator composed of $L$ and
$H$~\cite{Weinberg:1979sa}, one finds two more operators that involve
$N_R$~\cite{delAguila:2008ir,Aparici:2009fh}.

At $d=6$, in addition to the pure SMEFT operators~\cite{Grzadkowski:2010es},
there are five operators involving two fermions (at least one of which
is $N_R$) and bosons, eleven baryon and lepton-number-conserving 
four-fermion interactions, one lepton-number-violating operator,
and two operators that violate both baryon and lepton
number~\cite{Liao:2016qyd}.\footnote{Here, we count the operator
  types, \textit{i.e.}  we do not take into account the flavor
  structure and do not count hermitian conjugates.}  In the present
work, we are interested in the effects of the lepton-number-conserving four-fermion
interactions containing one $N_R$ and three SM fermions.  We list them
in Table~\ref{tab:singleNops}.  The effects of the four-fermion
operators containing a pair of HNLs and a pair of quarks have been
investigated in detail in Ref.~\cite{Cottin:2021lzz}.

\begin{table}[t]  
 \centering
 \renewcommand{\arraystretch}{1.2}
 \begin{tabular}[t]{|c|c|c|c|}
    \hline
    Name & Structure (+ h.c.) & $n_N = 1$ & $n_N = 3$ \\ 
\hline
\hline
${\cal O}_{duNe}$ &
$\left(\overline{d_R}\gamma^{\mu}u_R\right)\left(\overline{N_R}\gamma_{\mu}e_R\right)$ & 
   54  &
 162 \\ 
 ${\cal O}_{LNQd}$ &
 $\left(\overline{L}N_R\right)\epsilon\left(\overline{Q}d_R\right)$ &
  54 & 
  162 \\
  ${\cal O}_{LdQN}$ & 
  $\left(\overline{L}d_R\right) \epsilon \left(\overline{Q}N_R\right)$ &
   54 & 
   162 \\
 ${\cal O}_{QuNL}$ &
 $\left(\overline{Q}u_R\right)\left(\overline{N_R}L\right)$ &
  54 & 
  162 \\  
     \hline 
  ${\cal O}_{LNLe}$ &
  $\left(\overline{L}N_R\right) \epsilon \left(\overline{L}e_R\right)$ &
  54 & 
  162\\
\hline
 \end{tabular}
\caption{Lepton-number-conserving four-fermion single-$N_R$ operators.  For each operator
  structure, we provide the number of independent real parameters for
  $n_N = 1$ and $n_N= 3$ generations of $N_R$.  The operator in the
  last row is purely leptonic, and thus, it does not contribute to the
  HNL production at the LHC.}
\label{tab:singleNops}
\end{table}

The single-$N_R$ operators including quarks can lead to enhanced HNL
production cross section at the LHC, but they also trigger the decay
of $N_R$ to a lepton and two quarks.
The total decay width of the $N_R$'s depends on the operator.  Neglecting the masses of the lepton and light quarks, the partial decay width to charged
  leptons plus quarks is given by
\begin{equation}
 \Gamma(N_R \to \ell q q') =  \frac{c_\mathcal{O}^2 m_N^5}{f_\mathcal{O}512  \pi^3 \Lambda^4}\,,
 \label{eq:Gamma}
\end{equation}
with $m_N$ being the HNL mass, $c_\mathcal{O}$ the Wilson coefficient
of the operator $\mathcal{O}$, and $f_\mathcal{O}$ the numerical
factor depending on the operator type. For $\mathcal{O}_{duNe}$
$f_\mathcal{O} = 1$, whereas for $\mathcal{O}_{LNQd}$,
$\mathcal{O}_{LdQN}$, and $\mathcal{O}_{QuNL}$ $f_\mathcal{O} = 4$.
To arrive at the total decay width, one has to add also the final state
with neutrinos for all operators, except $\mathcal{O}_{duNe}$. Since
the partial width to neutrinos follows the same equation as for
charged leptons, this results in total decay widths being twice the partial decay widths given in Eq.~\eqref{eq:Gamma} (again, except for $\mathcal{O}_{duNe}$).
Finally, Eq.~\eqref{eq:Gamma} applies to Dirac neutrinos.
For Majorana neutrinos, one has to add also the charged conjugated channels, leading to another factor of 2 for the widths.

\subsection{Ultra-violet completions for four-fermion single-$N_R$ operators}
\label{sec:UVmodels}
%
The single-$N_R$ operators of interest can be generated in ultra-violet (UV) complete
models containing heavy scalars or vectors.  Here, we do not aim to
provide a complete classification of such UV completions, but rather give a few examples.  In what follows, we
consider scalar leptoquarks and an inert $SU(2)_L$ doublet scalar.  A
catalog of models with scalar and vector leptoquarks generating
four-fermion operators involving one or two $N_R$'s and quarks can be
found in Ref.~\cite{Bischer:2019ttk}.

The operator $\mathcal{O}_{duNe}$ can arise from a model with a scalar
leptoquark $S_d$ having the gauge quantum numbers of the down quark,
cf.~Table~\ref{tab:UV}.
\begin{table}[t]  
 \centering
 \renewcommand{\arraystretch}{1.5}
 \begin{tabular}{|l|ccc|c|c|}
\hline
Heavy scalar  & $SU(3)_C$ & $SU(2)_L$ & $U(1)_Y$ &  Operator & Matching relation  \\
\hline
\hline
Leptoquark $S_d$ & $\mathbf{3}$ & $\mathbf{1}$ & $-1/3$ &  ${\cal O}_{duNe}$ & $\dfrac{c_{duNe}}{\Lambda^2} = \dfrac{g_{dN} g_{ue}}{2 m_{S_d}^2}$ \\[0.3cm] 
\hline
Leptoquark $S_Q$ & $\mathbf{3}$ & $\mathbf{2}$ & $\phantom{-}1/6$ & ${\cal O}_{LdQN}$  &  $\dfrac{c_{LdQN}}{\Lambda^2}= \dfrac{g_{dL} g_{QN}}{m_{S_Q}^2}$ \\[0.3cm]
\hline 
\multirow{2.5}{*}{Inert doublet $\Phi$} & \multirow{2.5}{*}{$\mathbf{1}$} & \multirow{2.5}{*}{$\mathbf{2}$} & \multirow{2.5}{*}{$\phantom{-}1/2$} & ${\cal O}_{LNQd}$ & $\dfrac{c_{LNQd}}{\Lambda^2}= \dfrac{g_{LN} g_{Qd}}{m_{\Phi}^2}$ \\[0.3cm]
& & & & ${\cal O}_{QuNL}$ & $\dfrac{c_{QuNL}}{\Lambda^2}= \dfrac{g_{Qu} g_{LN}}{m_{\Phi}^2}$ \\[0.3cm] 
\hline
\end{tabular}
\caption{Heavy scalars with their gauge quantum numbers 
and the four-fermion single-$N_R$ operators they can generate. 
The last column reports the tree-level matching relations between 
the Wilson coefficients and the couplings of the UV model.}
\label{tab:UV}
\end{table}
The interaction Lagrangian of $S_d$ is given by
\begin{equation}
 -\mathcal{L}_{S_d} = g_{dN} \overline{d_R} N_R^c S_d + g_{ue} \overline{u_R} e_R^c S_d + g_{QL} \overline{Q} \epsilon L^c S_d + \text{h.c.}
 \label{eq:LSd}
\end{equation}
Upon integrating out $S_d$, the operator $\mathcal{O}_{duNe}$ is
generated with the tree-level matching condition for the Wilson
coefficient $c_{duNe}$ given in the last column of
Table~\ref{tab:UV}.\footnote{For simplicity, here, we assume the
  renormalizable couplings to be real and suppress flavor indices.
  The factor of two in the denominator originates from a Fierz
  identity.}  Analogously, a scalar leptoquark $S_Q$ with the quantum
numbers of the $SU(2)_L$ quark doublet can lead to
$\mathcal{O}_{LdQN}$.  The Yukawa interactions of $S_Q$ read
\begin{equation}
 -\mathcal{L}_{S_Q} = g_{QN} \overline{Q} N_R S_Q + g_{dL} \overline{d_R} L^T \epsilon S_Q + \text{h.c.}
 \label{eq:LSQ}
\end{equation}
We note that the first terms in Eqs.~\eqref{eq:LSd} and~\eqref{eq:LSQ}
also generate the $N_R$ pair operators $\mathcal{O}_{qN} =
(\overline{q} \gamma^\mu q) (\overline{N_R} \gamma_\mu N_R)$, where $q= d_R$ and $q=Q$, respectively, cf.~Ref.~\cite{Cottin:2021lzz}.

The operators $\mathcal{O}_{LNQd}$ and $\mathcal{O}_{QuNL}$, in turn,
can originate from a two Higgs doublet model, after the second, heavy
doublet $\Phi$ has been integrated out.  The interactions of interest
in the UV model have the following form:
\begin{equation}
 -\mathcal{L}_{\Phi} = g_{Qd} \overline{Q} \Phi d_R + g_{Qu} \overline{Q} \tilde{\Phi} u_R + g_{LN} \overline{L} \tilde{\Phi} N_R + \text{h.c.},
\end{equation}
where $\tilde{\Phi} = \epsilon \Phi^\ast$.  From Table~\ref{tab:UV},
it is clear that the Wilson coefficients of the operators depend on different combinations of independent
couplings in the UV model.  Therefore, in this example, the generated
operators are uncorrelated.

We have implemented these renormalizable models in \texttt{FeynRules}
\cite{Christensen:2008py,Alloul:2013bka} for both Dirac and Majorana
$N_R$.  Using the generated UFO~\cite{Degrande:2011ua} model files and
\texttt{MadGraph5}~\cite{Alwall:2011uj,Alwall:2014hca}, we have
checked that both cases lead to the same single-$N_R$ production cross
section.  We note that for $N_R$ pair production triggered by the
four-fermion operators with two $N_R$'s, the cross section is
different for Dirac and Majorana HNLs, especially for values of $m_N
\gtrsim 100$~GeV at LHC energies (we refer the interested reader
to Sec.~3.1 of Ref.~\cite{Cottin:2021lzz}).  The fact that the HNL
nature does not affect the production triggered by the four-fermion
single-$N_R$ operators allows us to implement these operators directly
in \texttt{FeynRules} for Dirac HNLs and use the resulting UFO model
file in \texttt{MadGraph5}.  (Recall that \texttt{MadGraph5} can
not handle Majorana fermions in operators with more than two fermions,
cf.~Sec.~3.1 of Ref.~\cite{Cottin:2021lzz}.)

%% file: tex/simulation.tex

\section{Simulation details\label{sect:sim}}

Our signal topology contains a prompt lepton and a displaced vertex
(DV) stemming from the $N_{R}$ decay to leptons and quarks.  Our stage
to reconstruct such a signature is the ATLAS detector, specifically
its inner tracker, as it has the capability to reconstruct vertices
displaced from the IP by few millimeters to tens of centimeters.  Our
analysis strategy builds up on an earlier work~\cite{Cottin:2018nms}
and is inspired from ATLAS multi-track displaced
searches~\cite{Aad:2015rba,Aaboud:2017iio}.

We consider the collision process $pp\to N l$ with $l=e,\mu,\tau$, at
$\sqrt{s}=14$ TeV at the high-luminosity LHC with an integrated
luminosity of 3 ab$^{-1}$.  We generate LHE events with displaced
information at the parton level with \texttt{MadGraph5}, which are
read by \texttt{Pythia8}~\cite{Sjostrand:2014zea} for showering and
hadronization. Our detector simulation is based on a custom made code
within \texttt{Pythia8}, where we first reconstruct isolated prompt
electrons, muons, and taus (with help from \texttt{FastJet}~\cite{Cacciari:2011ma}), taking into account detector acceptance,
resolution, and smearing on their transverse momenta (for details, see
Ref.~\cite{Cottin:2018nms}).
After selecting events with a prompt lepton, the displaced vertex
reconstruction starts by selecting tracks\footnote{A track in our
  simulation is a final state charged particle. These come from the
  decays of $N_{R}$ and can correspond to an electron, a muon, or a
  charged particle coming from the hadronization of quarks or from tau
  decays.} with $p_{T}>1$ GeV and a large impact parameter, $d_{0}$,
defined as $d_{0}=r_\text{trk} \times \Delta\phi$. Here $\Delta\phi$
corresponds to the azimuthal angle between the track and the direction
of the long-lived $N_{R}$, and $r_\text{trk}$ corresponds to the
transverse distance of the track from the origin. We require
$|d_{0}|>2$ mm. 

As we have access in simulation to truth-level Monte Carlo
information, we also identify the truth $N_{R}$ decay positions in the
transverse and longitudinal planes, namely, $r_{\text{DV}}$ and
$z_{\text{DV}}$, respectively.  An additional step (with respect to
Ref.~\cite{Cottin:2018nms}) of the vertex reconstruction implemented
in this work is the requirement that $r_\text{trk} - r_\text{DV} < 4$
mm.  It is not always the case that the ``starting" point of the
displaced track matches the displaced vertex position.  This is more
evident in the case where we have a tau produced from the $N_{R}$
displaced decay, as taus also have an additional
displacement.\footnote{The proper decay distance of tau leptons is
  $c\tau=87.1$ $\mu$m. This will lead, for example, to decay distances
  of $\gamma c\tau\sim 5$mm at 100 GeV.}  With this requirement, we
emulate what an experimental displaced-vertex reconstruction would do
when fitting nearby displaced tracks to a common
origin~\cite{Aad:2015rba}.  This will lead to an additional reduction
in efficiency when reconstructing displaced vertices containing taus.
Nevertheless, it is a more realistic (and optimistic) approach than
what was done in Ref.~\cite{Cottin:2018nms} to handle heavy neutrino
decays to taus (see Sec.~\ref{sect:res} below).

After selecting optimal displaced tracks, we demand displaced vertices
within the ATLAS inner tracker acceptance, namely, $4$ mm
$<r_{\text{DV}}<300$ mm and $|z_{\text{DV}}|<300$ mm.  Further cuts
are applied on the number of high-quality tracks coming from the DV,
$N_\text{trk}$, and its invariant mass, $m_{\text{DV}}$, assuming all
tracks have the pion mass.  More concretely, we require
$N_\text{trk}>3$ and $m_{\text{DV}} \geq 5$ GeV.  As detailed in
Refs.~\cite{Aad:2015rba,Aaboud:2017iio,Cottin:2018kmq}, these last two
cuts ensure that we are in a region where signal is expected to be
found free of backgrounds including $B$-mesons.  Further detector
response to DVs is quantified by applying the 13 TeV ATLAS
parameterized efficiencies~\cite{Aaboud:2017iio} as a function of DV
invariant mass and number of tracks, where we assume these will remain
the same at 14 TeV.

%% file: tex/results.tex

\section{Numerical results\label{sect:res}}

Based on the computational procedure described in the previous
section, we have estimated the experimental sensitivities (95 $\%$
confidence level (C.L.) exclusion limits under the assumption of zero
background) of searches for long-lived HNLs at the ATLAS detector for
two different theoretical scenarios. The first is the minimal scenario
in which only right-handed neutrinos, $N_R$, are added to the particle
content of the SM and renormalizable interactions are assumed.  In
this case, the HNLs interact with the SM particles only through the
mixing with the active neutrinos, $V_{lN}$, with $ l = e, \mu, \tau $.
In the second theoretical scenario we consider $N_R$SMEFT containing
non-renormalizable interactions of $N_R$ with the SM.  In this case,
both production and decay of the HNLs can be mediated by the
single-$N_R$ effective operators under consideration.  In all of the
plots below, we assume a $3+1$ scenario where the HNL mixes dominantly
with only one active neutrino flavor at a time. We assume only one HNL
is kinematically relevant.

\subsection{Minimal scenario}
%
In the minimal scenario, the relevant parameters are the mass of the
HNL, $m_N$, and the mixing of the HNL with the active neutrinos,
$V_{lN}$, which we have treated as independent parameters.  The HNLs
are produced from the decays of on-shell $W$-bosons into a lepton and
an HNL associated with a charged lepton, $ pp \rightarrow W
\rightarrow l N $, via the HNL mixing with the active neutrinos.  The
decay of the HNLs occurs also via the mixing with the active
neutrinos, through both charged and neutral SM currents,
$ N \rightarrow l (\nu) jj $. For the minimal scenario we use the
\texttt{FeynRules} implementation for HNLs of
Ref.~\cite{Degrande:2016aje}.

\begin{figure}[t]
	\centering
	\includegraphics[width=0.49\textwidth]{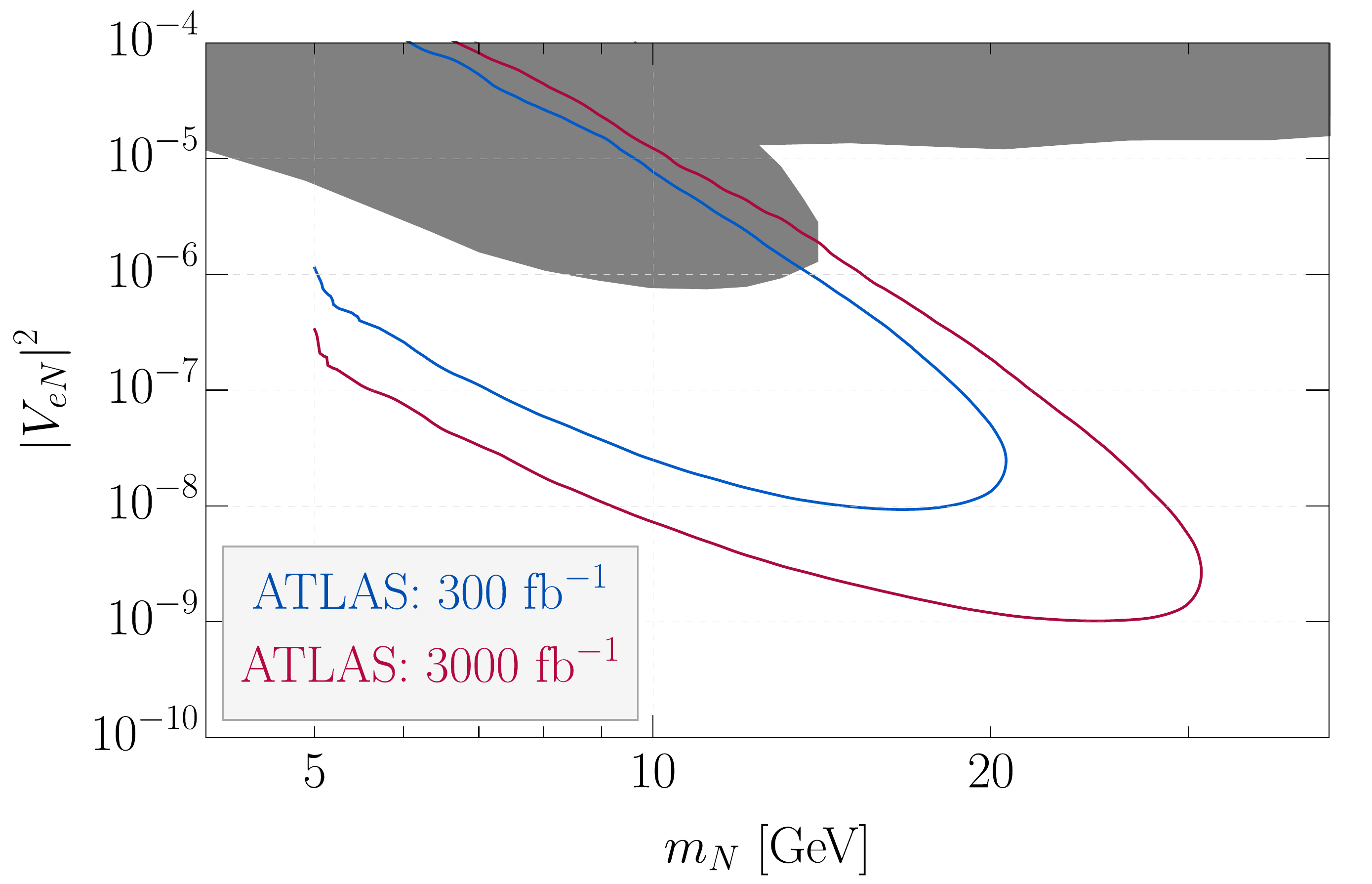}
	\includegraphics[width=0.49\textwidth]{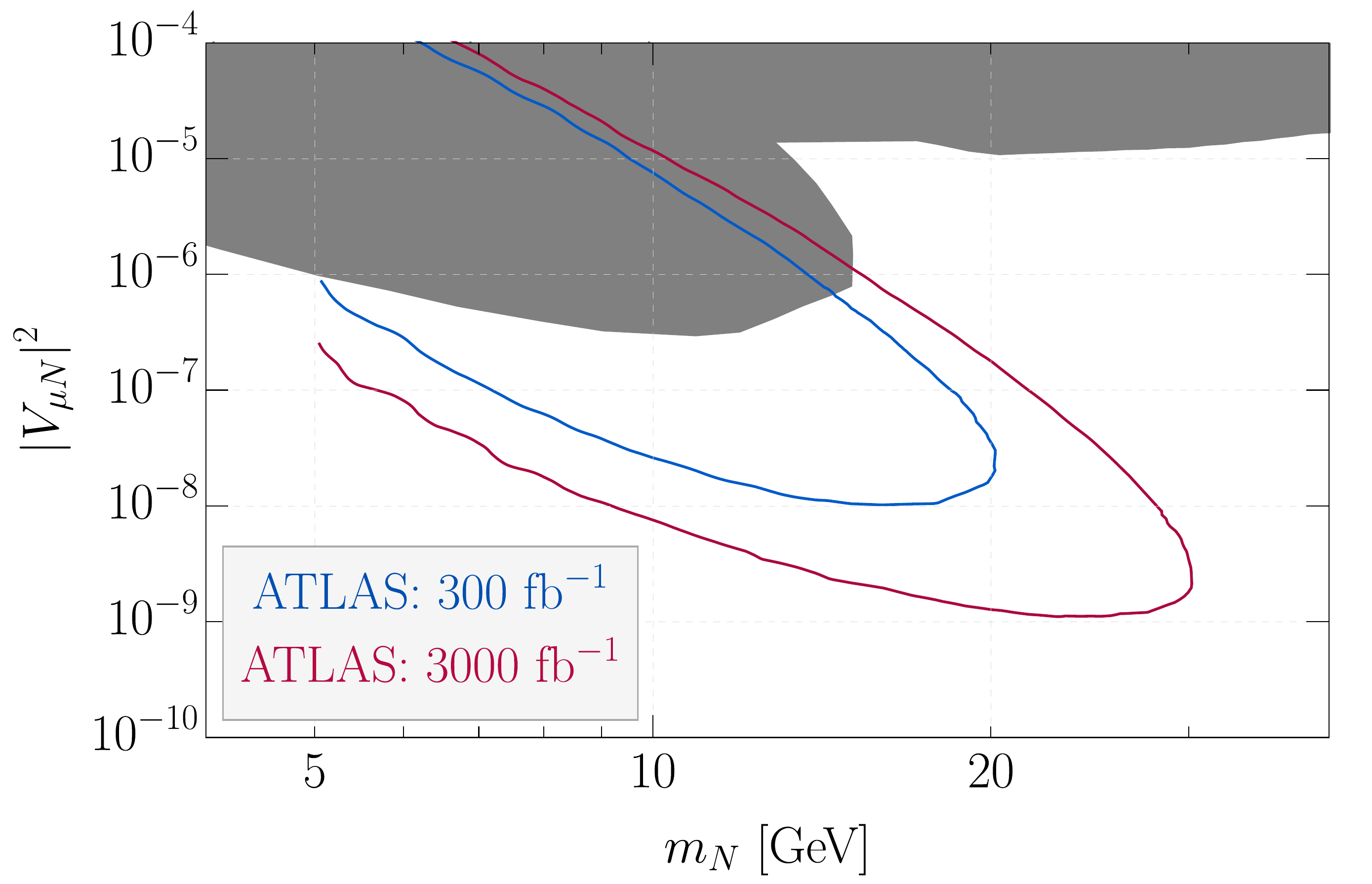}\\
	\includegraphics[width=0.49\textwidth]{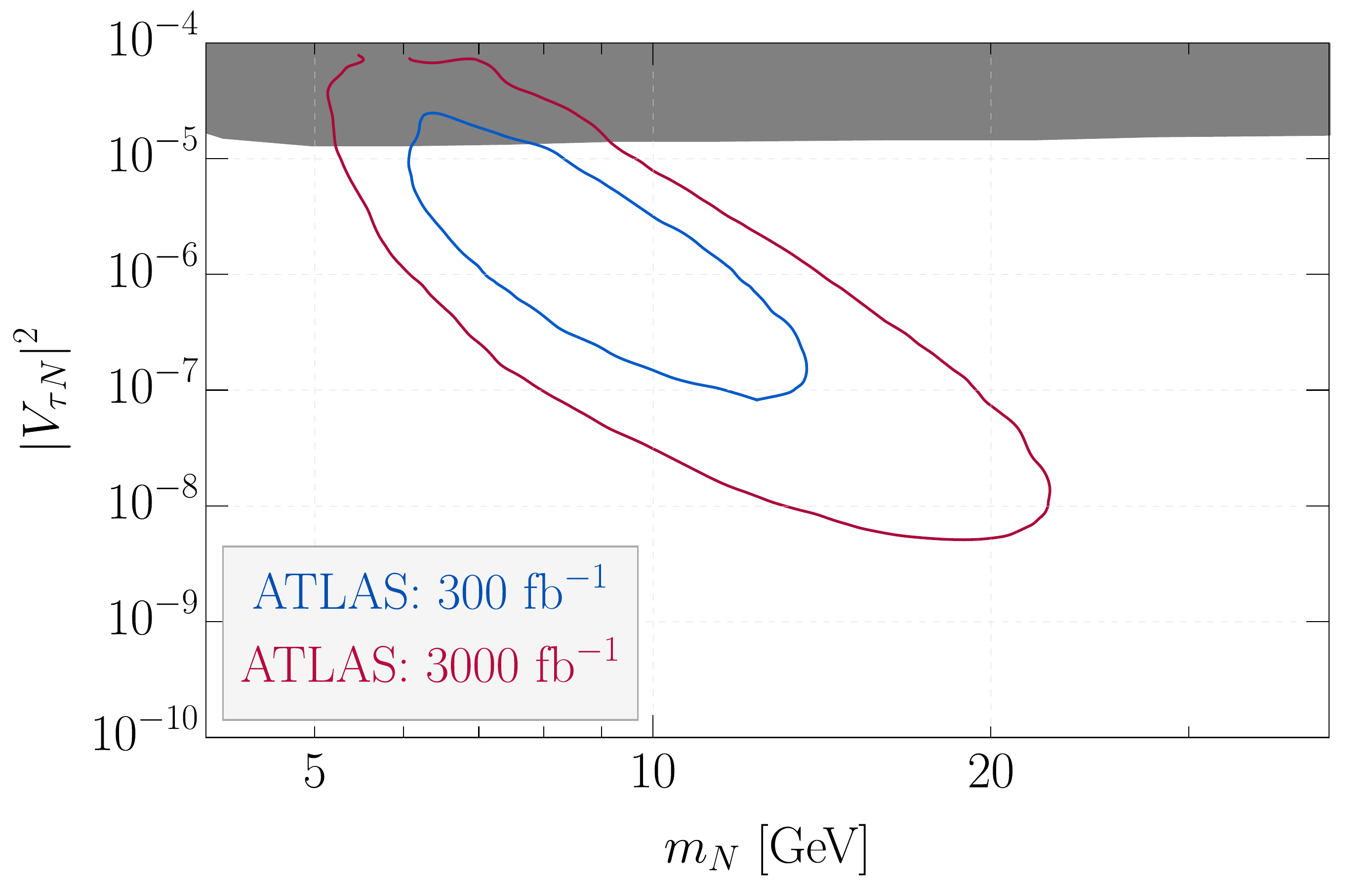}
	\caption{Minimal scenario sensitivity reach on $|V_{lN}|^2$ as
          a function of $m_N$, for $l= e, \mu, \tau$. The dark region
          corresponds to current experimental limits obtained at
          several experiments: ATLAS~\cite{ATLAS:2019kpx},
          CMS~\cite{CMS:2018iaf,CMS:2021lzm}, DELPHI~\cite{Abreu:1996pa}, and
          LHCb~\cite{LHCb:2016inz,Antusch:2017hhu}.
        } \label{fig:minimalsensitivity}
\end{figure}

Figure~\ref{fig:minimalsensitivity} shows the region, in the plane
$|V_{lN}|^2$ vs.~$m_N$, where a displaced-vertex search at the ATLAS
detector for the center-of-mass energy 14 TeV, and with the selection
criteria discussed in Sec.~\ref{sect:sim}, may have sensitivity to the
minimal scenario.    As can be
seen in this figure, the sensitivities in $|V_{eN}|^2$ and
$|V_{\mu N}|^2$ are rather similar and can reach values down to
$|V_{lN}|^2 \sim 10^{-9}$ for $m_N \sim 30$ GeV, with $3$ ab$^{-1}$ of
integrated luminosity.  On the other hand, in the case of mixing with
the tau neutrinos, ATLAS can reach values of the mixing parameter down
to $|V_{\tau N}|^2 \sim 5 \times 10^{-9}$ for $ m_N \sim 20$ GeV with
$3$ ab$^{-1}$.  Figure~\ref{fig:minimalsensitivity} compares our
limits with the current experimental bounds for this model,
represented by the dark gray area at the top of each plot.  These
constraints were obtained at the following experiments:
ATLAS~\cite{ATLAS:2019kpx}, CMS~\cite{CMS:2018iaf,CMS:2021lzm},
DELPHI~\cite{Abreu:1996pa}, and LHCb~\cite{LHCb:2016inz,Antusch:2017hhu}.
As we can see, our forecast limits can reach values of the mixing
$|V_{lN}|^2$ several orders of magnitude smaller than current
experimental bounds.

It might be interesting to compare the forecast limits to theory
expectations. In seesaw type-I, one naively expects $|V_{lN}|^2 \simeq
m_{\nu}/m_N \simeq (10^{-12}-10^{-11})$ for values of $m_N$ in the range
we are considering in this work. Larger values of $|V_{lN}|^2$ are
possible allowing fine-tuning in parameters. A more natural model
for having $|V_{lN}|^2$ in the range accessible for ATLAS/CMS might
be the inverse seesaw \cite{Mohapatra:1986bd}. In this variant
of the seesaw, $|V_{lN}|^2$ is given by $|V_{lN}|^2 \simeq m_{\nu}/\mu$,
with $\mu$ being the lepton-number-violating parameter of the
inverse seesaw model.  With $\mu$ supposedly being a small parameter,
when compared to $m_N$, mixings in this model can be easily as large as the
experimental limits. 

As mentioned above, the same search strategy for long-lived HNLs was
previously proposed by some of us in Ref.~\cite{Cottin:2018nms}.  One
of the differences between our current and previous calculations is
the center-of-mass energy at the LHC, which now is taken as 14~TeV
(previously in Ref.~\cite{Cottin:2018nms} we used 13~TeV).  Perhaps
more important is the fact that in the present paper, our numerical
calculations used more statistics, which allowed us to obtain much
smoother contours for our limits, which led to a slight increase in
the ranges shown.  Moreover, in the case of mixing with taus, our
current limits are more sensitive than the previous ones calculated in
Ref.~\cite{Cottin:2018nms}.  The reason for this difference is that in
Ref.~\cite{Cottin:2018nms}, we only considered neutral currents in the
decay of HNLs that coupled to taus (i.e. we ignored a tau lepton
coming from the displaced vertex), whereas now, we have included both
charged and neutral currents in our calculations, making our limits
more realistic for the case of the mixing with the tau neutrinos and
comparable with the sensitivity reach projected with other proposed
strategies (see for instance Ref.~\cite{Drewes:2019fou}).

\subsection{Four-fermion single-$N_R$ operators}
%
In the second theoretical scenario, we consider the four-fermion
single-$N_R$ operators in the $N_R$SMEFT.  We estimate the
experimental sensitivity of our displaced search to a long-lived HNL
at the ATLAS detector.  Here, we take the coefficients of the
operators $c_{{\cal O}}/\Lambda^2$ and the mass of the HNL, $m_N$, as
independent parameters.  In this scenario, both the production and the
decay of the HNL can be dominated by the same operator $ {\cal O}$,
unlike the case of effective operators with two
HNLs~\cite{Cottin:2021lzz}, where the pair-$N_R$ operators dominantly
induce the HNL production, but the decay of the HNL still proceeds
only via mixing with the active neutrinos.  For the EFT scenario, in
our analysis we have assumed that the contributions to the production
and decay of the HNL from its mixing with active neutrinos $ V_{lN} $
are sub-dominant and negligible compared to the effective operators'
contributions. For mixing angles smaller than $|V_{lN}|^2 \lesssim
10^{-9}$, this assumption is always fulfilled.

The production of the HNLs considered in our analysis, $ p p
\rightarrow l N $, is always accompanied by a prompt charged
lepton~---~ an electron, muon, or tau, depending on the flavor
structure of the effective operator considered.  The presence of this
charged lepton is important in our analysis as it is used to trigger
the signal, as discussed in Sec.~\ref{sect:sim}.  The decay of the
HNLs will occur via the same operator leading to two jets and one
neutral or charged lepton, $ N \rightarrow l (\nu) j j $.  The
production cross sections of the HNLs will depend on the type of
quarks that the respective operator includes.  In our analysis, we
have only considered effective operators with quarks of the first two
generations.

\begin{figure}[t]
	\centering
	\includegraphics[width=0.49\textwidth]{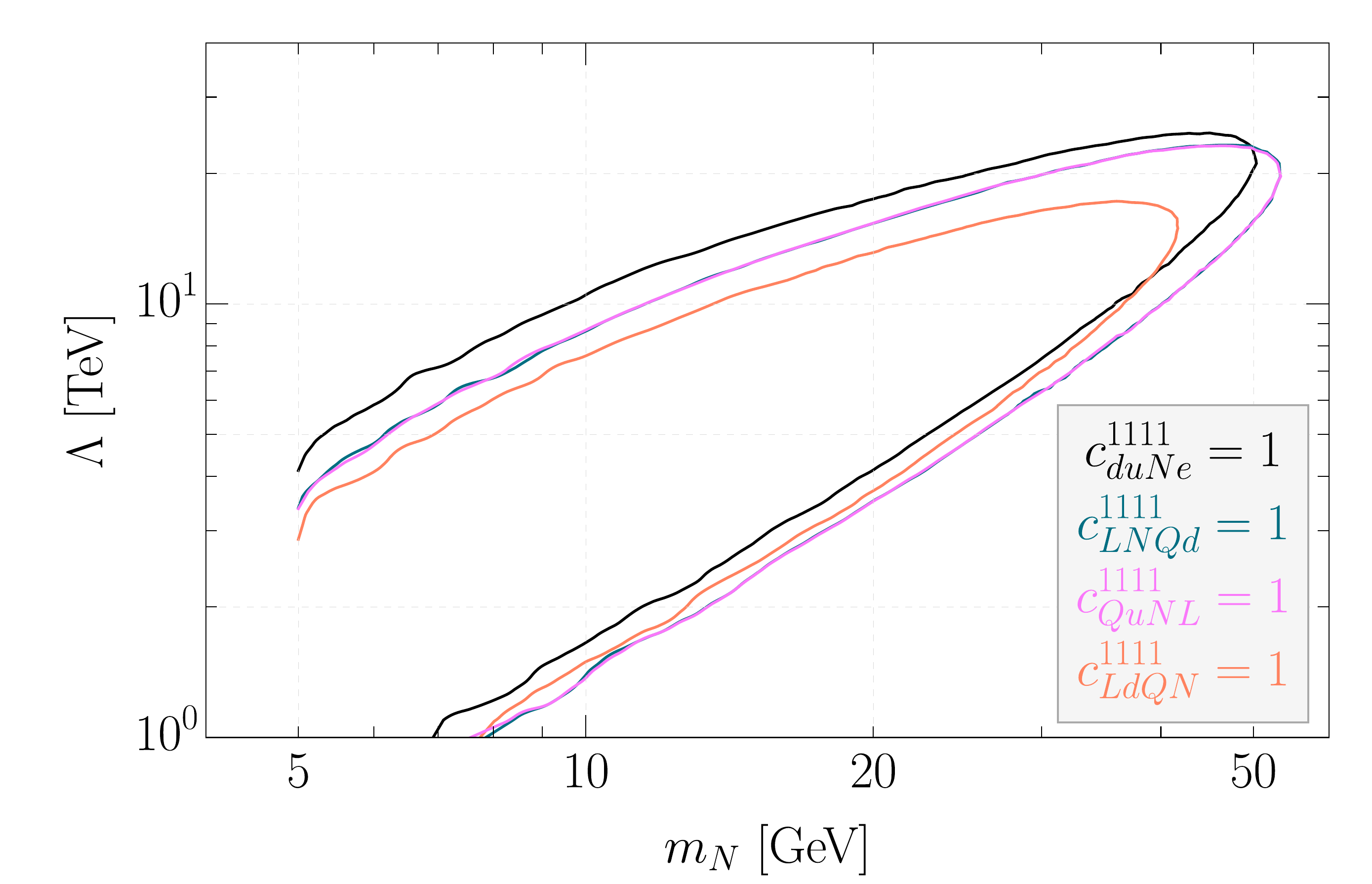}
	\includegraphics[width=0.49\textwidth]{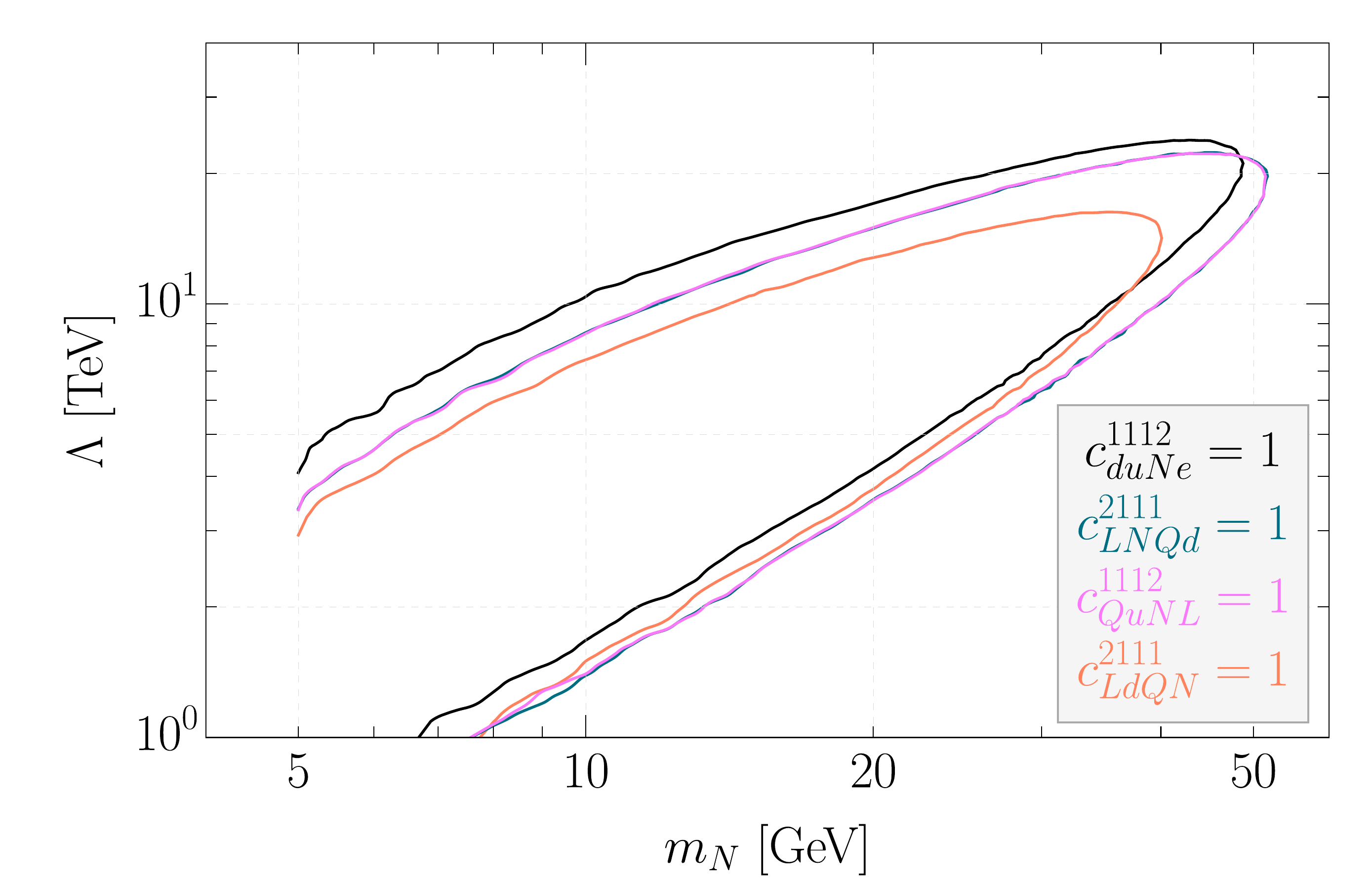}\\
	\includegraphics[width=0.49\textwidth]{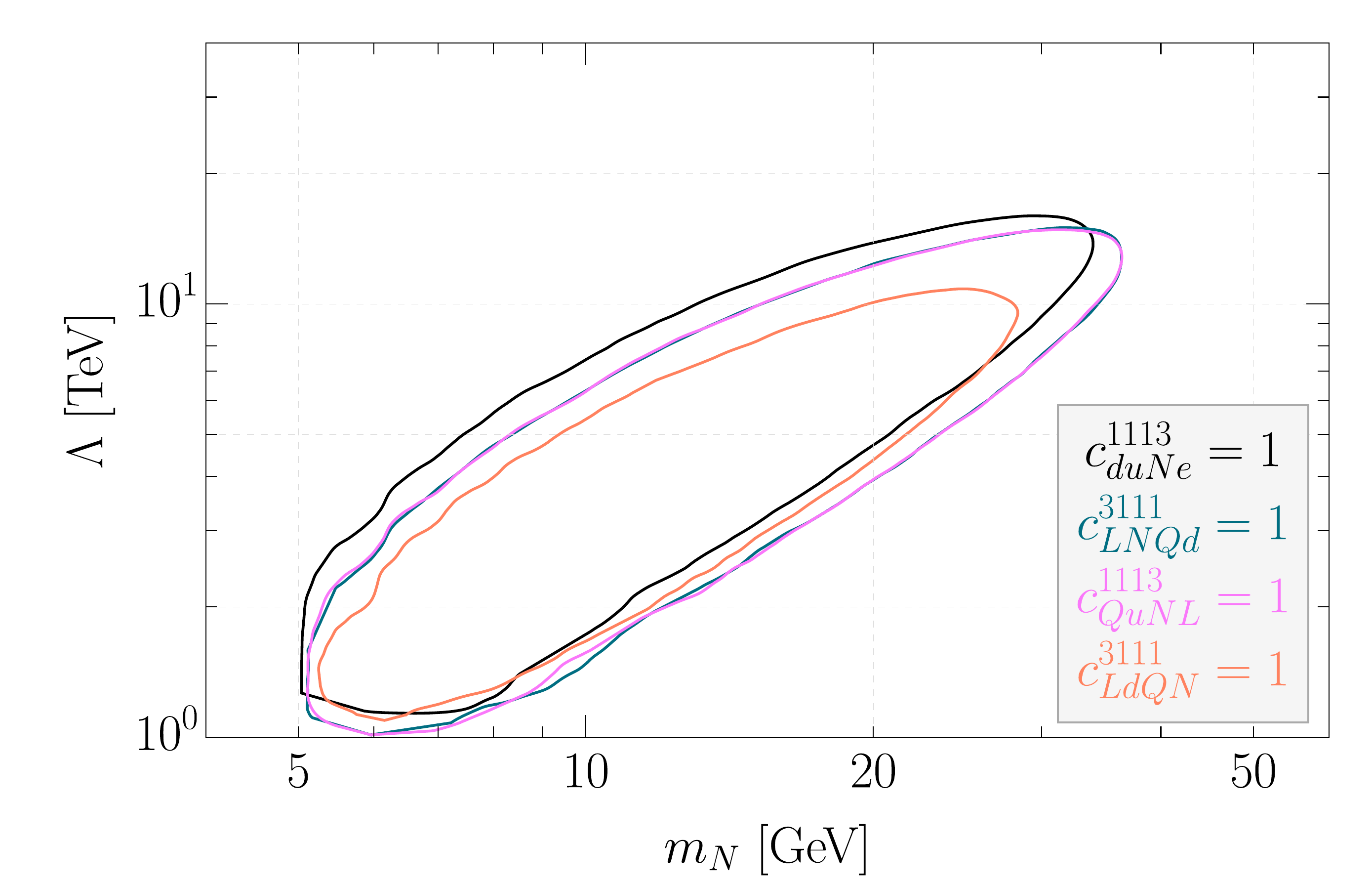}
	\caption{Exclusion limits on the new physics scale $\Lambda$
          as a function of $m_N$ in the EFT scenario with operators
          including the first-generation quarks only, for an
          integrated luminosity of 3 ab$^{-1}$.  The two plots at the
          top consider operators with charged leptons of the first and
          second generation: electrons (left) and muons (right). The
          plot at the bottom considers operators with tau leptons
          only.} \label{fig:eftsensitivity}
\end{figure}

In Fig.~\ref{fig:eftsensitivity}, we show the experimental sensitivity
of the ATLAS detector to a long-lived HNL in the $\Lambda$ vs.~$m_N$
plane.  In our analysis, we have considered the contributions of one
operator at a time, setting the value of the corresponding operator
coefficient $c_{{\cal O}} = 1$, and the rest of the operator
coefficients to zero.  In Fig.~\ref{fig:eftsensitivity}, we have
considered only operators with quarks of the first generation.  Note
that the numbers in the superscript of \textit{e.g.}
$c_{duNe}^{1112}$ refer to the first-generation quarks ($d$ and
$u$), the lightest $N_R$ and the second-generation charged lepton
(the muon).  As can be seen in this figure, for an integrated
luminosity of 3~ab$^{-1}$, ATLAS can reach values of the new physics
scale up to (and above) $\Lambda \sim 20$ TeV for masses $m_N
\lsim 50$ GeV in the case of operators with an electron or muon.  In
the case of operators with a tau lepton, ATLAS can reach $\Lambda$
$\gtrsim 10$ TeV at masses $m_N$ of 10's GeV.  It is worth mentioning
that our limits start at $ m_N \gtrsim 5$ GeV.  The reason is the
kinematic cut at $ m_\text{DV}\geq  5$ GeV imposed in the selection
criteria.  This cut is necessary to remove the SM background coming
from $B$-mesons, as discussed in Sec.~\ref{sect:sim}.
We also note that the projected exclusion limits are rather similar 
for the four types of single-$N_R$ operators, in particular for 
$\mathcal{O}_{LNQd}$ and $\mathcal{O}_{QuNL}$.

\begin{figure}[t]
	\centering
	\includegraphics[width=0.49\textwidth]{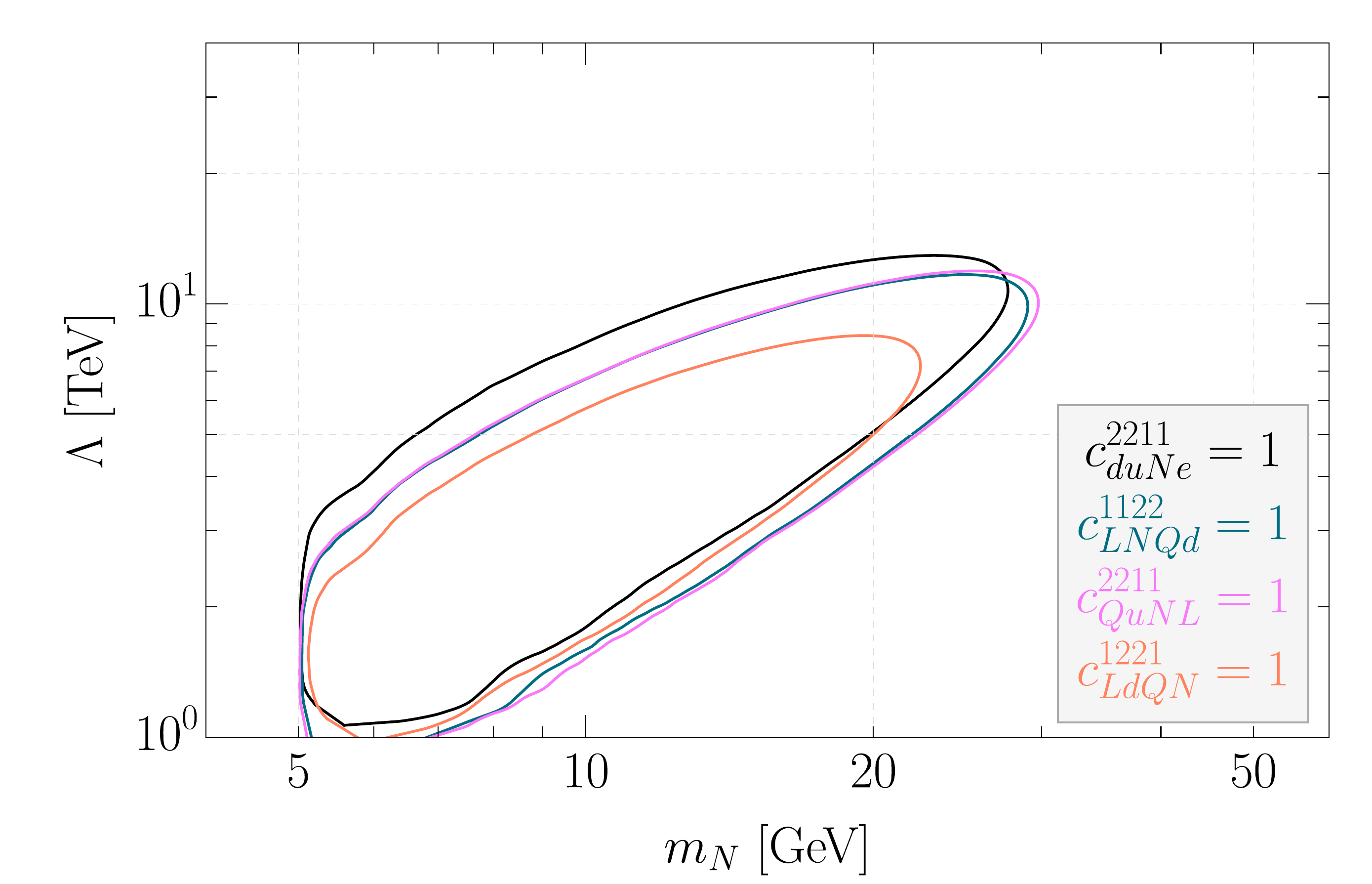}
	\includegraphics[width=0.49\textwidth]{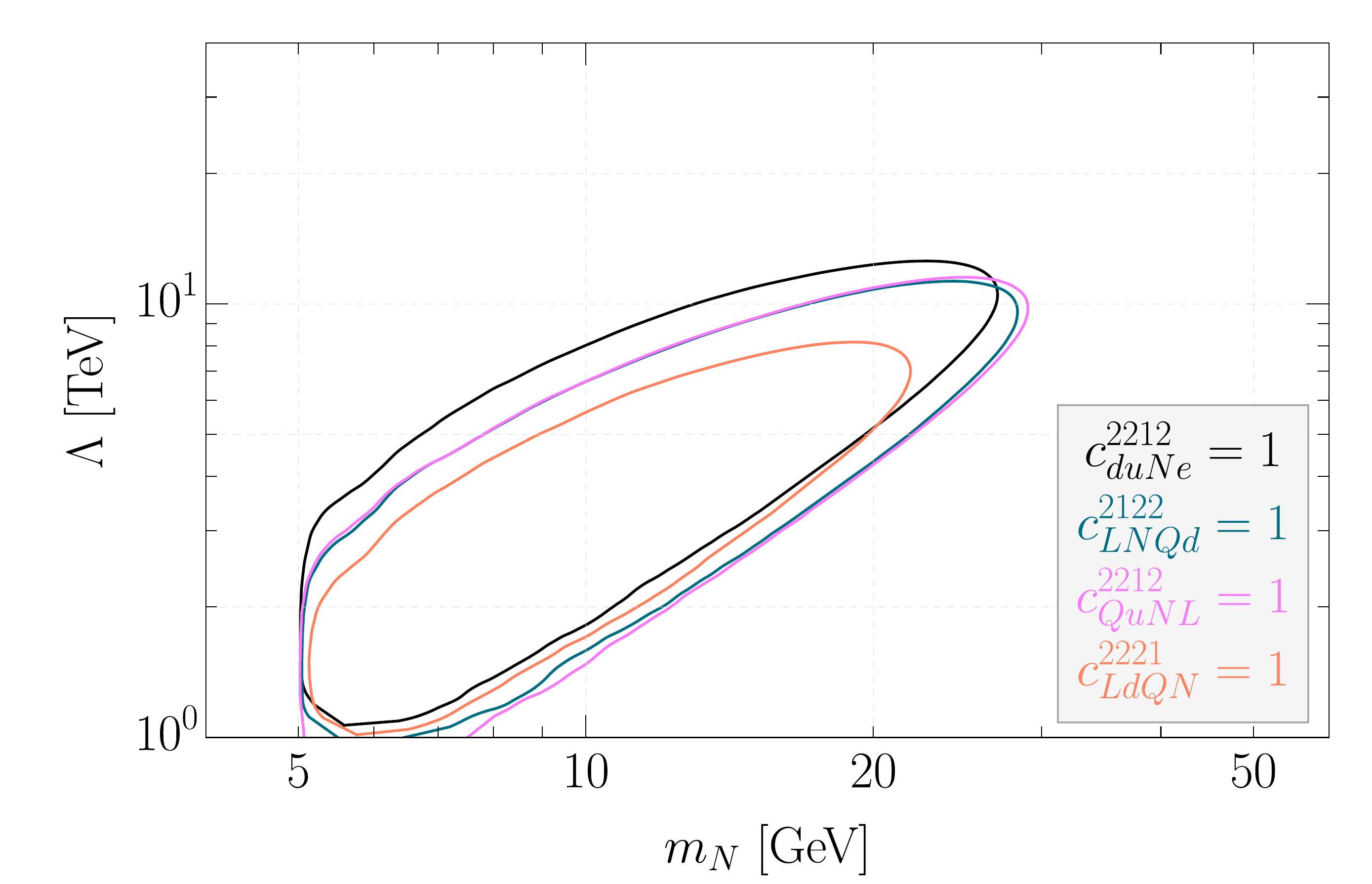}\\
	\includegraphics[width=0.49\textwidth]{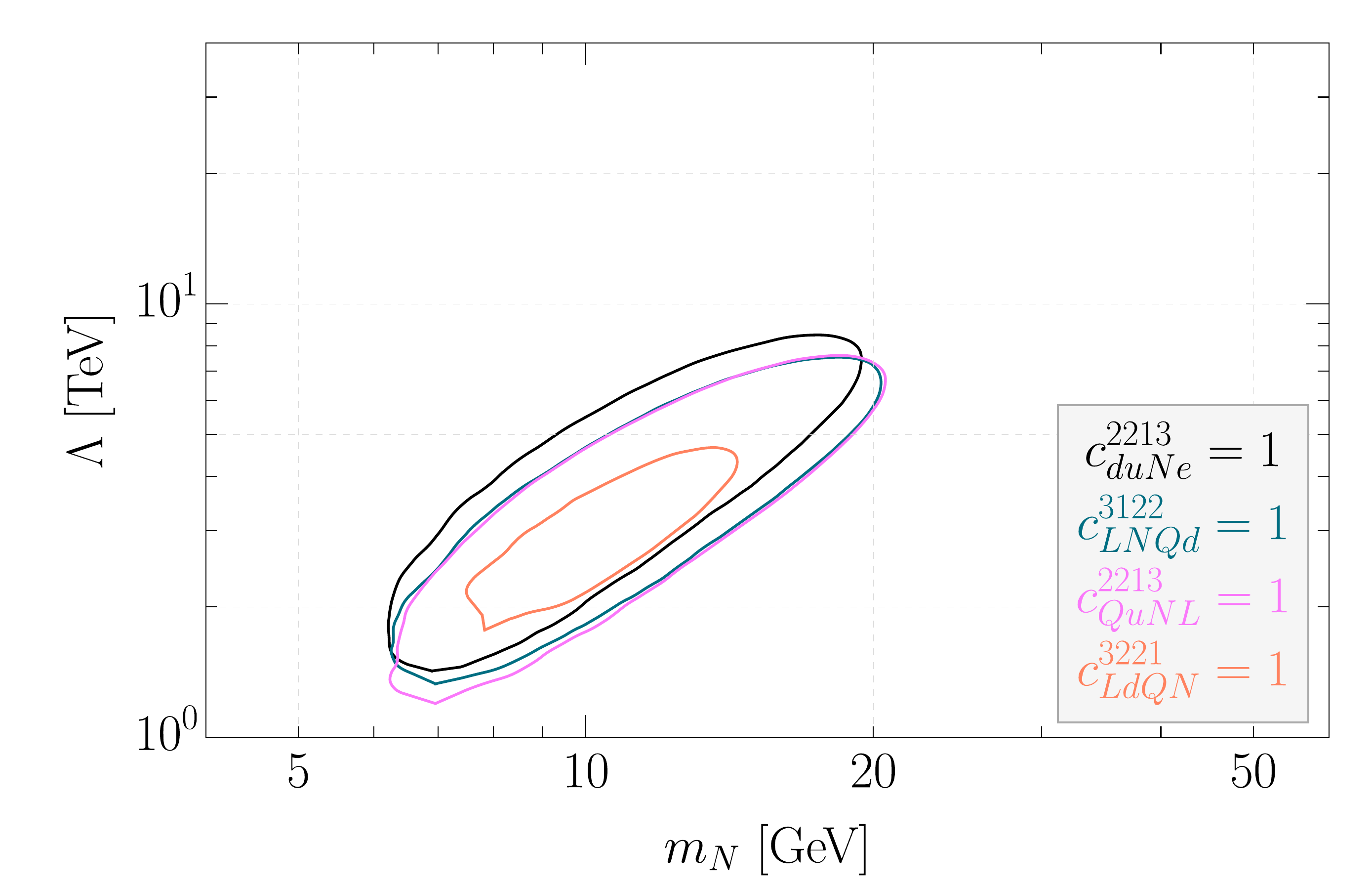}
	\caption{The same as Fig.~\ref{fig:eftsensitivity}, but for
          operators with second-generation quarks
          only.} \label{fig:eftsensitivity2}
\end{figure}

Figure~\ref{fig:eftsensitivity2} contains our limits in the plane
$\Lambda$ vs.~$m_N$ for the effective operators with quarks of the
second generation only.  As expected, the sensitivity regions for
operators with quarks of the second generation only are smaller than
those corresponding to operators with first-generation quarks
(Fig.~\ref{fig:eftsensitivity}).  This is due to the predominant
content of quarks $u$ and $d$ in the proton versus the quarks $c$ and
$s$.  We find that limits shown in Fig.~\ref{fig:eftsensitivity2} can
reach $ \Lambda \sim 13$ TeV for $m_N \sim 23 $ GeV in the cases of
electrons and muons, and up to $\Lambda \sim 9 $ TeV for
$ m_N \sim 18$ GeV in the case of taus. All numbers assume an
integrated luminosity of 3 ab$^{-1}$.  Other possible combinations of
quark flavors for the $N_R$SMEFT include $(u,s)$ and $(c,d)$.  The
sensitivity reaches for these operators lie between the two cases
shown in Figs.~\ref{fig:eftsensitivity} and
\ref{fig:eftsensitivity2} (for $(u,d)$ and $(c,s)$). We therefore do
not show results for these cases explicitly.  Operators with 
third-generation quarks have not been considered in this work, since they
will require special treatment (i.e. tagging).

We also note that Figs.~\ref{fig:eftsensitivity} and \ref{fig:eftsensitivity2} have been calculated for Dirac HNLs.  As
mentioned above, production cross sections for single-$N_R$ operators
are the same for Dirac and Majorana HNLs, while the half-lives for
Majorana HNLs are smaller by a factor of two. The sensitivity regions
for Majorana HNLs therefore differ slightly from the regions shown in
the figures. We do not repeat the plots for the Majorana case and
instead opt for a short explanation of the differences. First, the
maximal value of the HNL mass, to which this kind of search is
sensitive is determined by the smallest decay length that is
accessible in the experiment. Since the decay width scales as
$m_N^5$, for a Majorana HNL the largest HNL mass accessible is a
factor $(1/2)^{1/5} \simeq 0.87$ smaller than in the Dirac case.
Second, the maximal value of $\Lambda$ reached in our sensitivity
curves is essentially determined by the total cross section (times
luminosity). Since cross sections are the same for Dirac and Majorana
HNLs, this maximal value of $\Lambda$ does not change for Majorana
HNLs. Finally, in the regime where the decay lengths are large (i.e.
for large values of $\Lambda$ at values of $m_{N}$ smaller than the
one where the maximal value of $\Lambda$ is reached), the event number
depends linearly on the half-life, while both the cross section and the decay width scale as $\Lambda^{-4}$. For Majorana HNLs, in
this part of the parameter space, slightly larger values of $\Lambda$
are accessible than for the Dirac case, i.e. an increase by
roughly a factor $2^{1/8} \simeq 1.09$.

Let us briefly comment on existing limits. First of all, our choice of
switching ``on'' always only one Wilson coefficient at a time
guarantees that there is no new source of lepton (or quark) flavour
violation. It is well known that ultra-violet completions, such as the
leptoquark models we discuss in Sec.~\ref{sec:UVmodels}, are strongly
constrained by searches for lepton-flavor-violating processes.  These
will put lower bounds on $m_{LQ}\simeq\Lambda$ that are much stronger
than anything achievable in direct searches
\cite{Davidson:1993qk,PDG2020}.  Thus, {\em in all accelerator
  searches} it is customary to assume that new resonances, such as
leptoquarks, couple only to one SM fermion generation at a
time. Direct searches for leptoquarks from pair and single leptoquark
production have been performed by both, CMS
\cite{CMS:2018ncu,CMS:2018yiq,CMS:2019ybf} and ATLAS
\cite{ATLAS:2016wab,ATLAS:2020dsk}. The best limits approach now
$m_{LQ} \simeq 2$ TeV. Thus, the long-lived particle search discussed
in this paper, will probe so-far uncharted parts of parameter
space. Let us also mention, that since $d=6$ operators have Wilson
coefficients of the form $c/\Lambda^2$, limits on $\Lambda$ will scale
proportional to $\sqrt{c}$. Thus for $c \lsim 10^{-2}$ our search will
no longer probe values of $\Lambda$ not already excluded by direct LHC
searches.

We will close this discussion with one additional comment. Our
simulated analysis focused on the ATLAS detector and its
reconstruction capabilities to displaced vertices inside the inner
tracker, starting from 4 mm in multi-track
searches~\cite{Aad:2015rba,Aaboud:2017iio}.  Relaxing this requirement
to decay distances below 4 mm (both in $d_{0}$, $r_{\text{DV}}$ and
$z_{\text{DV}}$) will allow to extend the reach in parameter space
towards larger HNL masses.  Of course, with the loosening of these
cuts we may depart from the zero background case assumption, and a
detailed study on the multi-track search backgrounds would be needed,
which goes beyond the scope of the present work. Nevertheless, past
displaced lepton searches -- whose tracks are fitted to a common
vertex -- at CMS~\cite{CMS:2014hka} could probe transverse decay
lengths starting from $\approx 200$~$\mu$m\footnote{The explicit
  analysis requirement in~Ref.~\cite{CMS:2014hka} demands tracks to
  have a transverse impact parameter significance with respect to the
  primary vertex of $|d_{0}|/\sigma_{d}>12$, where $\sigma_{d}$ is the
  uncertainty on $|d_{0}|$. }.  In addition, a recent 13 TeV CMS
search~\cite{CMS:2021kdm} demonstrates that lepton tracks with
$|d_{0}|>0.1$ mm are displaced enough to be considered for analysis.
Finally, the recent CMS note on an HNL search with an explicit
displaced vertex requirement does not even demand a constraint on the
DV minimal distance~\cite{CMS:2021lzm}.  This provides feasibility to
experimentally go below the 4 mm threshold.  We stress that an
improvement of the displaced vertex search towards smaller decay
lengths by such a larger factor (up to $40$ for $0.1$mm) would allow
to test HNL masses larger by a factor 2 w.r.t the values in our
figures, i.e. extend the searches from $m_{N} \simeq 50$ GeV to
roughly $100$ GeV.  We hope that this large potential gain motivates
the experimental collaborations to study the lowering of the
transverse cuts in displaced vertex searches to the sub-millimeter
range.

%% file: tex/summary.tex

\section{Summary}\label{sect:sum}

The Standard Model (SM) effective field theory (EFT) extended with
sterile neutrinos, also known as the $N_R$SMEFT, provides a framework
to systematically study sterile neutrinos associated with a high
new-physics (NP) scale in ultra-violet complete models beyond the SM.
In the $N_R$SMEFT, high-scale NP effects are encoded in the so-called
Wilson coefficients of non-renormalizable operators at different mass
dimensions.  Higher-dimensional operators involving $N_R$ can have
either one, two, or four sterile neutrinos, and may conserve or
violate lepton number, or else both lepton and baryon numbers.

In this work, we have focused on lepton-number-conserving four-fermion
single-$N_R$ operators associated with a charged lepton and two
quarks, which can induce both production and decay of the heavy
neutral leptons (HNLs) simultaneously.  For HNLs of
$\mathcal{O}(10)$~GeV mass, such operators with a NP scale above $\sim
1$ TeV can easily make the HNLs become long-lived, leading to
displaced vertices at the LHC.  We have therefore proposed a
displaced-vertex search strategy based on a prompt-lepton trigger and
selection of high-quality displaced tracks.  By performing Monte-Carlo
simulations with \texttt{MadGraph}5 and \texttt{Pythia}8, we have
estimated the sensitivity reaches for ATLAS in the high-luminosity LHC
era with 3 ab$^{-1}$ integrated luminosity, to four single-$N_R$ EFT
operators: $\mathcal{O}_{duNe}$, $\mathcal{O}_{LNQd}$,
$\mathcal{O}_{LdQN}$, and $\mathcal{O}_{QuNL}$.

Multiple combinations of quark and lepton flavors can be studied.
Here, we have considered mainly two combinations: $(u,d)$ and $(c,s)$.
The first (second)-generation-quark only flavor combination is then
projected to have the best (worst) sensitivities, because of their
portion in the proton content.  For both quark combinations, we also
studied all possible lepton generations, i.e. electron, muon and tau.
In addition, for simplicity, we did not take into account the effect
of the active-sterile neutrino mixing, which is supposed to be
negligible if the type-I seesaw relation is assumed.  For the $(u,d)$
and $(c,s)$ combinations, we find in general for the considered
single-$N_R$ operators, ATLAS can probe $\Lambda$ up to 20 TeV
and above for $m_N\gtrsim 20$ GeV, if we switch on one
operator at a time.

In addition to the EFT scenarios, we also revisited the minimal
scenario of the HNL mixing with the SM neutrinos. In this scenario,
the type-I seesaw relation is not assumed and we have two independent
parameters: mass of the HNL and its mixing parameter with one type of
the active neutrinos: a simple $3+1$ scenario.  These results are an
update of those given in Ref.~\cite{Cottin:2018nms}.  Besides some
minor changes, the most important difference is that we have now taken
into account both charged and neutral currents in our computation,
leading to more realistic projection results, especially for the case
of mixing with the $\tau$ neutrino.

In summary, we conclude that a displaced-vertex search at ATLAS for
HNLs can probe new physics scales up to about $20$~TeV and, in some
cases above, for HNL mass between about 5 GeV and 50 GeV, depending on
the quark and lepton flavors associated with the single-$N_R$ operator
under consideration.

%% file: SingleN_EFT.bbl
\providecommand{\href}[2]{#2}\begingroup\raggedright\begin{thebibliography}{10}

\bibitem{Alimena:2019zri}
J.~Alimena et~al., \emph{{Searching for long-lived particles beyond the
  Standard Model at the Large Hadron Collider}},
  \href{https://doi.org/10.1088/1361-6471/ab4574}{\emph{J. Phys. G} {\bfseries
  47} (2020) 090501} [\href{https://arxiv.org/abs/1903.04497}{{\ttfamily
  1903.04497}}].

\bibitem{Lee:2018pag}
L.~Lee, C.~Ohm, A.~Soffer and T.-T.~Yu, \emph{{Collider Searches for Long-Lived
  Particles Beyond the Standard Model}},
  \href{https://doi.org/10.1016/j.ppnp.2019.02.006}{\emph{Prog. Part. Nucl.
  Phys.} {\bfseries 106} (2019) 210}
  [\href{https://arxiv.org/abs/1810.12602}{{\ttfamily 1810.12602}}].

\bibitem{Curtin:2018mvb}
D.~Curtin et~al., \emph{{Long-Lived Particles at the Energy Frontier: The
  MATHUSLA Physics Case}},
  \href{https://doi.org/10.1088/1361-6633/ab28d6}{\emph{Rept. Prog. Phys.}
  {\bfseries 82} (2019) 116201}
  [\href{https://arxiv.org/abs/1806.07396}{{\ttfamily 1806.07396}}].

\bibitem{Minkowski:1977sc}
P.~Minkowski, \emph{{$\mu \to e\gamma$ at a Rate of One Out of $10^{9}$ Muon
  Decays?}}, \href{https://doi.org/10.1016/0370-2693(77)90435-X}{\emph{Phys.
  Lett. B} {\bfseries 67} (1977) 421}.

\bibitem{Yanagida:1979as}
T.~Yanagida, \emph{{Horizontal gauge symmetry and masses of neutrinos}},
  {\emph{Conf. Proc. C} {\bfseries 7902131} (1979) 95}.

\bibitem{GellMann:1980vs}
M.~Gell-Mann, P.~Ramond and R.~Slansky, \emph{{Complex Spinors and Unified
  Theories}}, {\emph{Conf. Proc. C} {\bfseries 790927} (1979) 315}
  [\href{https://arxiv.org/abs/1306.4669}{{\ttfamily 1306.4669}}].

\bibitem{Mohapatra:1979ia}
R.N.~Mohapatra and G.~Senjanovic, \emph{{Neutrino Mass and Spontaneous Parity
  Nonconservation}},
  \href{https://doi.org/10.1103/PhysRevLett.44.912}{\emph{Phys. Rev. Lett.}
  {\bfseries 44} (1980) 912}.

\bibitem{Schechter:1980gr}
J.~Schechter and J.W.F.~Valle, \emph{{Neutrino Masses in SU(2) x U(1)
  Theories}}, \href{https://doi.org/10.1103/PhysRevD.22.2227}{\emph{Phys. Rev.
  D} {\bfseries 22} (1980) 2227}.

\bibitem{deSalas:2020pgw}
P.F.~de~Salas, D.V.~Forero, S.~Gariazzo, P.~Mart\'\i{}nez-Mirav\'e, O.~Mena,
  C.A.~Ternes et~al., \emph{{2020 global reassessment of the neutrino
  oscillation picture}},
  \href{https://doi.org/10.1007/JHEP02(2021)071}{\emph{JHEP} {\bfseries 02}
  (2021) 071} [\href{https://arxiv.org/abs/2006.11237}{{\ttfamily
  2006.11237}}].

\bibitem{Capozzi:2021fjo}
F.~Capozzi, E.~Di~Valentino, E.~Lisi, A.~Marrone, A.~Melchiorri and A.~Palazzo,
  \emph{{The unfinished fabric of the three neutrino paradigm}},
  \href{https://arxiv.org/abs/2107.00532}{{\ttfamily 2107.00532}}.

\bibitem{Esteban:2020cvm}
I.~Esteban, M.C.~Gonzalez-Garcia, M.~Maltoni, T.~Schwetz and A.~Zhou,
  \emph{{The fate of hints: updated global analysis of three-flavor neutrino
  oscillations}}, \href{https://doi.org/10.1007/JHEP09(2020)178}{\emph{JHEP}
  {\bfseries 09} (2020) 178}
  [\href{https://arxiv.org/abs/2007.14792}{{\ttfamily 2007.14792}}].

\bibitem{Mohapatra:1986bd}
R.~Mohapatra and J.~Valle, \emph{{Neutrino Mass and Baryon Number
  Nonconservation in Superstring Models}},
  \href{https://doi.org/10.1103/PhysRevD.34.1642}{\emph{Phys. Rev.} {\bfseries
  D34} (1986) 1642}.

\bibitem{Bernabeu:1987gr}
J.~Bernabeu, A.~Santamaria, J.~Vidal, A.~Mendez and J.W.F.~Valle, \emph{{Lepton
  Flavor Nonconservation at High-Energies in a Superstring Inspired Standard
  Model}}, \href{https://doi.org/10.1016/0370-2693(87)91100-2}{\emph{Phys.
  Lett. B} {\bfseries 187} (1987) 303}.

\bibitem{Akhmedov:1995ip}
E.K.~Akhmedov, M.~Lindner, E.~Schnapka and J.~Valle, \emph{{Left-right symmetry
  breaking in NJL approach}},
  \href{https://doi.org/10.1016/0370-2693(95)01504-3}{\emph{Phys.Lett.}
  {\bfseries B368} (1996) 270}
  [\href{https://arxiv.org/abs/hep-ph/9507275}{{\ttfamily hep-ph/9507275}}].

\bibitem{Akhmedov:1995vm}
E.K.~Akhmedov, M.~Lindner, E.~Schnapka and J.~Valle, \emph{{Dynamical
  left-right symmetry breaking}},
  \href{https://doi.org/10.1103/PhysRevD.53.2752}{\emph{Phys.Rev.} {\bfseries
  D53} (1996) 2752} [\href{https://arxiv.org/abs/hep-ph/9509255}{{\ttfamily
  hep-ph/9509255}}].

\bibitem{Mohapatra:1974hk}
R.N.~Mohapatra and J.C.~Pati, \emph{{Left-Right Gauge Symmetry and an
  Isoconjugate Model of CP Violation}},
  \href{https://doi.org/10.1103/PhysRevD.11.566}{\emph{Phys. Rev. D} {\bfseries
  11} (1975) 566}.

\bibitem{Senjanovic:1975rk}
G.~Senjanovic and R.N.~Mohapatra, \emph{{Exact Left-Right Symmetry and
  Spontaneous Violation of Parity}},
  \href{https://doi.org/10.1103/PhysRevD.12.1502}{\emph{Phys. Rev. D}
  {\bfseries 12} (1975) 1502}.

\bibitem{Brivio:2017vri}
I.~Brivio and M.~Trott, \emph{{The Standard Model as an Effective Field
  Theory}}, \href{https://doi.org/10.1016/j.physrep.2018.11.002}{\emph{Phys.
  Rept.} {\bfseries 793} (2019) 1}
  [\href{https://arxiv.org/abs/1706.08945}{{\ttfamily 1706.08945}}].

\bibitem{Ellis:2020unq}
J.~Ellis, M.~Madigan, K.~Mimasu, V.~Sanz and T.~You, \emph{{Top, Higgs, Diboson
  and Electroweak Fit to the Standard Model Effective Field Theory}},
  \href{https://doi.org/10.1007/JHEP04(2021)279}{\emph{JHEP} {\bfseries 04}
  (2021) 279} [\href{https://arxiv.org/abs/2012.02779}{{\ttfamily
  2012.02779}}].

\bibitem{Ethier:2021bye}
J.J.~Ethier, G.~Magni, F.~Maltoni, L.~Mantani, E.R.~Nocera, J.~Rojo et~al.,
  \emph{{Combined SMEFT interpretation of Higgs, diboson, and top quark data
  from the LHC}},  \href{https://arxiv.org/abs/2105.00006}{{\ttfamily
  2105.00006}}.

\bibitem{delAguila:2008ir}
F.~del Aguila, S.~Bar-Shalom, A.~Soni and J.~Wudka, \emph{{Heavy Majorana
  Neutrinos in the Effective Lagrangian Description: Application to Hadron
  Colliders}},
  \href{https://doi.org/10.1016/j.physletb.2008.11.031}{\emph{Phys. Lett. B}
  {\bfseries 670} (2009) 399}
  [\href{https://arxiv.org/abs/0806.0876}{{\ttfamily 0806.0876}}].

\bibitem{Aparici:2009fh}
A.~Aparici, K.~Kim, A.~Santamaria and J.~Wudka, \emph{{Right-handed neutrino
  magnetic moments}},
  \href{https://doi.org/10.1103/PhysRevD.80.013010}{\emph{Phys. Rev. D}
  {\bfseries 80} (2009) 013010}
  [\href{https://arxiv.org/abs/0904.3244}{{\ttfamily 0904.3244}}].

\bibitem{Bhattacharya:2015vja}
S.~Bhattacharya and J.~Wudka, \emph{{Dimension-seven operators in the standard
  model with right handed neutrinos}},
  \href{https://doi.org/10.1103/PhysRevD.94.055022}{\emph{Phys. Rev. D}
  {\bfseries 94} (2016) 055022}
  [\href{https://arxiv.org/abs/1505.05264}{{\ttfamily 1505.05264}}].

\bibitem{Liao:2016qyd}
Y.~Liao and X.-D.~Ma, \emph{{Operators up to Dimension Seven in Standard Model
  Effective Field Theory Extended with Sterile Neutrinos}},
  \href{https://doi.org/10.1103/PhysRevD.96.015012}{\emph{Phys. Rev. D}
  {\bfseries 96} (2017) 015012}
  [\href{https://arxiv.org/abs/1612.04527}{{\ttfamily 1612.04527}}].

\bibitem{Li:2021tsq}
H.-L.~Li, Z.~Ren, M.-L.~Xiao, J.-H.~Yu and Y.-H.~Zheng, \emph{{Operator Bases
  in Effective Field Theories with Sterile Neutrinos: $d \leq 9$}},
  \href{https://arxiv.org/abs/2105.09329}{{\ttfamily 2105.09329}}.

\bibitem{Chala:2020vqp}
M.~Chala and A.~Titov, \emph{{One-loop matching in the SMEFT extended with a
  sterile neutrino}},
  \href{https://doi.org/10.1007/JHEP05(2020)139}{\emph{JHEP} {\bfseries 05}
  (2020) 139} [\href{https://arxiv.org/abs/2001.07732}{{\ttfamily
  2001.07732}}].

\bibitem{Chala:2020pbn}
M.~Chala and A.~Titov, \emph{{One-loop running of dimension-six Higgs-neutrino
  operators and implications of a large neutrino dipole moment}},
  \href{https://doi.org/10.1007/JHEP09(2020)188}{\emph{JHEP} {\bfseries 09}
  (2020) 188} [\href{https://arxiv.org/abs/2006.14596}{{\ttfamily
  2006.14596}}].

\bibitem{Datta:2020ocb}
A.~Datta, J.~Kumar, H.~Liu and D.~Marfatia, \emph{{Anomalous dimensions from
  gauge couplings in SMEFT with right-handed neutrinos}},
  \href{https://doi.org/10.1007/JHEP02(2021)015}{\emph{JHEP} {\bfseries 02}
  (2021) 015} [\href{https://arxiv.org/abs/2010.12109}{{\ttfamily
  2010.12109}}].

\bibitem{Datta:2021akg}
A.~Datta, J.~Kumar, H.~Liu and D.~Marfatia, \emph{{Anomalous dimensions from
  Yukawa couplings in SMNEFT: four-fermion operators}},
  \href{https://doi.org/10.1007/JHEP05(2021)037}{\emph{JHEP} {\bfseries 05}
  (2021) 037} [\href{https://arxiv.org/abs/2103.04441}{{\ttfamily
  2103.04441}}].

\bibitem{Bischer:2019ttk}
I.~Bischer and W.~Rodejohann, \emph{{General neutrino interactions from an
  effective field theory perspective}},
  \href{https://doi.org/10.1016/j.nuclphysb.2019.114746}{\emph{Nucl. Phys. B}
  {\bfseries 947} (2019) 114746}
  [\href{https://arxiv.org/abs/1905.08699}{{\ttfamily 1905.08699}}].

\bibitem{Alcaide:2019pnf}
J.~Alcaide, S.~Banerjee, M.~Chala and A.~Titov, \emph{{Probes of the Standard
  Model effective field theory extended with a right-handed neutrino}},
  \href{https://doi.org/10.1007/JHEP08(2019)031}{\emph{JHEP} {\bfseries 08}
  (2019) 031} [\href{https://arxiv.org/abs/1905.11375}{{\ttfamily
  1905.11375}}].

\bibitem{Butterworth:2019iff}
J.M.~Butterworth, M.~Chala, C.~Englert, M.~Spannowsky and A.~Titov,
  \emph{{Higgs phenomenology as a probe of sterile neutrinos}},
  \href{https://doi.org/10.1103/PhysRevD.100.115019}{\emph{Phys. Rev. D}
  {\bfseries 100} (2019) 115019}
  [\href{https://arxiv.org/abs/1909.04665}{{\ttfamily 1909.04665}}].

\bibitem{Biekotter:2020tbd}
A.~Biek\"otter, M.~Chala and M.~Spannowsky, \emph{{The effective field theory
  of low scale see-saw at colliders}},
  \href{https://doi.org/10.1140/s10052-020-8339-2}{\emph{Eur. Phys. J. C}
  {\bfseries 80} (2020) 743}
  [\href{https://arxiv.org/abs/2007.00673}{{\ttfamily 2007.00673}}].

\bibitem{Dekens:2020ttz}
W.~Dekens, J.~de~Vries, K.~Fuyuto, E.~Mereghetti and G.~Zhou, \emph{{Sterile
  neutrinos and neutrinoless double beta decay in effective field theory}},
  \href{https://doi.org/10.1007/JHEP06(2020)097}{\emph{JHEP} {\bfseries 06}
  (2020) 097} [\href{https://arxiv.org/abs/2002.07182}{{\ttfamily
  2002.07182}}].

\bibitem{Han:2020pff}
T.~Han, J.~Liao, H.~Liu and D.~Marfatia, \emph{{Scalar and tensor neutrino
  interactions}}, \href{https://doi.org/10.1007/JHEP07(2020)207}{\emph{JHEP}
  {\bfseries 07} (2020) 207}
  [\href{https://arxiv.org/abs/2004.13869}{{\ttfamily 2004.13869}}].

\bibitem{Li:2020lba}
T.~Li, X.-D.~Ma and M.A.~Schmidt, \emph{{General neutrino interactions with
  sterile neutrinos in light of coherent neutrino-nucleus scattering and meson
  invisible decays}},
  \href{https://doi.org/10.1007/JHEP07(2020)152}{\emph{JHEP} {\bfseries 07}
  (2020) 152} [\href{https://arxiv.org/abs/2005.01543}{{\ttfamily
  2005.01543}}].

\bibitem{Li:2020wxi}
T.~Li, X.-D.~Ma and M.A.~Schmidt, \emph{{Constraints on the charged currents in
  general neutrino interactions with sterile neutrinos}},
  \href{https://doi.org/10.1007/JHEP10(2020)115}{\emph{JHEP} {\bfseries 10}
  (2020) 115} [\href{https://arxiv.org/abs/2007.15408}{{\ttfamily
  2007.15408}}].

\bibitem{DeVries:2020jbs}
J.~De~Vries, H.K.~Dreiner, J.Y.~G\"unther, Z.S.~Wang and G.~Zhou,
  \emph{{Long-lived Sterile Neutrinos at the LHC in Effective Field Theory}},
  \href{https://doi.org/10.1007/JHEP03(2021)148}{\emph{JHEP} {\bfseries 03}
  (2021) 148} [\href{https://arxiv.org/abs/2010.07305}{{\ttfamily
  2010.07305}}].

\bibitem{Cottin:2021lzz}
G.~Cottin, J.C.~Helo, M.~Hirsch, A.~Titov and Z.S.~Wang, \emph{{Heavy neutral
  leptons in effective field theory and the high-luminosity LHC}},
  \href{https://doi.org/10.1007/JHEP09(2021)039}{\emph{JHEP} {\bfseries 09}
  (2021) 039} [\href{https://arxiv.org/abs/2105.13851}{{\ttfamily
  2105.13851}}].

\bibitem{Caputo:2017pit}
A.~Caputo, P.~Hernandez, J.~Lopez-Pavon and J.~Salvado, \emph{{The seesaw
  portal in testable models of neutrino masses}},
  \href{https://doi.org/10.1007/JHEP06(2017)112}{\emph{JHEP} {\bfseries 06}
  (2017) 112} [\href{https://arxiv.org/abs/1704.08721}{{\ttfamily
  1704.08721}}].

\bibitem{Barducci:2020icf}
D.~Barducci, E.~Bertuzzo, A.~Caputo, P.~Hernandez and B.~Mele, \emph{{The
  see-saw portal at future Higgs Factories}},
  \href{https://doi.org/10.1007/JHEP03(2021)117}{\emph{JHEP} {\bfseries 03}
  (2021) 117} [\href{https://arxiv.org/abs/2011.04725}{{\ttfamily
  2011.04725}}].

\bibitem{Chou:2016lxi}
J.P.~Chou, D.~Curtin and H.J.~Lubatti, \emph{{New Detectors to Explore the
  Lifetime Frontier}},
  \href{https://doi.org/10.1016/j.physletb.2017.01.043}{\emph{Phys. Lett. B}
  {\bfseries 767} (2017) 29}
  [\href{https://arxiv.org/abs/1606.06298}{{\ttfamily 1606.06298}}].

\bibitem{Alpigiani:2020tva}
{\scshape MATHUSLA} collaboration, \emph{{An Update to the Letter of Intent for
  MATHUSLA: Search for Long-Lived Particles at the HL-LHC}},
  \href{https://arxiv.org/abs/2009.01693}{{\ttfamily 2009.01693}}.

\bibitem{Gligorov:2017nwh}
V.V.~Gligorov, S.~Knapen, M.~Papucci and D.J.~Robinson, \emph{{Searching for
  Long-lived Particles: A Compact Detector for Exotics at LHCb}},
  \href{https://doi.org/10.1103/PhysRevD.97.015023}{\emph{Phys. Rev. D}
  {\bfseries 97} (2018) 015023}
  [\href{https://arxiv.org/abs/1708.09395}{{\ttfamily 1708.09395}}].

\bibitem{Gligorov:2018vkc}
V.V.~Gligorov, S.~Knapen, B.~Nachman, M.~Papucci and D.J.~Robinson,
  \emph{{Leveraging the ALICE/L3 cavern for long-lived particle searches}},
  \href{https://doi.org/10.1103/PhysRevD.99.015023}{\emph{Phys. Rev. D}
  {\bfseries 99} (2019) 015023}
  [\href{https://arxiv.org/abs/1810.03636}{{\ttfamily 1810.03636}}].

\bibitem{Feng:2017uoz}
J.L.~Feng, I.~Galon, F.~Kling and S.~Trojanowski, \emph{{ForwArd Search
  ExpeRiment at the LHC}},
  \href{https://doi.org/10.1103/PhysRevD.97.035001}{\emph{Phys. Rev. D}
  {\bfseries 97} (2018) 035001}
  [\href{https://arxiv.org/abs/1708.09389}{{\ttfamily 1708.09389}}].

\bibitem{Bauer:2019vqk}
M.~Bauer, O.~Brandt, L.~Lee and C.~Ohm, \emph{{ANUBIS: Proposal to search for
  long-lived neutral particles in CERN service shafts}},
  \href{https://arxiv.org/abs/1909.13022}{{\ttfamily 1909.13022}}.

\bibitem{Helo:2018qej}
J.C.~Helo, M.~Hirsch and Z.S.~Wang, \emph{{Heavy neutral fermions at the
  high-luminosity LHC}},
  \href{https://doi.org/10.1007/JHEP07(2018)056}{\emph{JHEP} {\bfseries 07}
  (2018) 056} [\href{https://arxiv.org/abs/1803.02212}{{\ttfamily
  1803.02212}}].

\bibitem{Dercks:2018wum}
D.~Dercks, H.K.~Dreiner, M.~Hirsch and Z.S.~Wang, \emph{{Long-Lived Fermions at
  AL3X}}, \href{https://doi.org/10.1103/PhysRevD.99.055020}{\emph{Phys. Rev. D}
  {\bfseries 99} (2019) 055020}
  [\href{https://arxiv.org/abs/1811.01995}{{\ttfamily 1811.01995}}].

\bibitem{Hirsch:2020klk}
M.~Hirsch and Z.S.~Wang, \emph{{Heavy neutral leptons at ANUBIS}},
  \href{https://doi.org/10.1103/PhysRevD.101.055034}{\emph{Phys. Rev. D}
  {\bfseries 101} (2020) 055034}
  [\href{https://arxiv.org/abs/2001.04750}{{\ttfamily 2001.04750}}].

\bibitem{Zhou:2021ylt}
G.~Zhou, J.Y.~G\"unther, Z.S.~Wang, J.~de~Vries and H.K.~Dreiner,
  \emph{{Long-lived Sterile Neutrinos at Belle II in Effective Field Theory}},
  \href{https://arxiv.org/abs/2111.04403}{{\ttfamily 2111.04403}}.

\bibitem{Cottin:2018nms}
G.~Cottin, J.C.~Helo and M.~Hirsch, \emph{{Displaced vertices as probes of
  sterile neutrino mixing at the LHC}},
  \href{https://doi.org/10.1103/PhysRevD.98.035012}{\emph{Phys. Rev. D}
  {\bfseries 98} (2018) 035012}
  [\href{https://arxiv.org/abs/1806.05191}{{\ttfamily 1806.05191}}].

\bibitem{Weinberg:1979sa}
S.~Weinberg, \emph{{Baryon and Lepton Nonconserving Processes}},
  \href{https://doi.org/10.1103/PhysRevLett.43.1566}{\emph{Phys. Rev. Lett.}
  {\bfseries 43} (1979) 1566}.

\bibitem{Grzadkowski:2010es}
B.~Grzadkowski, M.~Iskrzynski, M.~Misiak and J.~Rosiek, \emph{{Dimension-Six
  Terms in the Standard Model Lagrangian}},
  \href{https://doi.org/10.1007/JHEP10(2010)085}{\emph{JHEP} {\bfseries 10}
  (2010) 085} [\href{https://arxiv.org/abs/1008.4884}{{\ttfamily 1008.4884}}].

\bibitem{Christensen:2008py}
N.D.~Christensen and C.~Duhr, \emph{{FeynRules - Feynman rules made easy}},
  \href{https://doi.org/10.1016/j.cpc.2009.02.018}{\emph{Comput. Phys. Commun.}
  {\bfseries 180} (2009) 1614}
  [\href{https://arxiv.org/abs/0806.4194}{{\ttfamily 0806.4194}}].

\bibitem{Alloul:2013bka}
A.~Alloul, N.D.~Christensen, C.~Degrande, C.~Duhr and B.~Fuks, \emph{{FeynRules
  2.0 - A complete toolbox for tree-level phenomenology}},
  \href{https://doi.org/10.1016/j.cpc.2014.04.012}{\emph{Comput. Phys. Commun.}
  {\bfseries 185} (2014) 2250}
  [\href{https://arxiv.org/abs/1310.1921}{{\ttfamily 1310.1921}}].

\bibitem{Degrande:2011ua}
C.~Degrande, C.~Duhr, B.~Fuks, D.~Grellscheid, O.~Mattelaer and T.~Reiter,
  \emph{{UFO - The Universal FeynRules Output}},
  \href{https://doi.org/10.1016/j.cpc.2012.01.022}{\emph{Comput. Phys. Commun.}
  {\bfseries 183} (2012) 1201}
  [\href{https://arxiv.org/abs/1108.2040}{{\ttfamily 1108.2040}}].

\bibitem{Alwall:2011uj}
J.~Alwall, M.~Herquet, F.~Maltoni, O.~Mattelaer and T.~Stelzer, \emph{{MadGraph
  5 : Going Beyond}},
  \href{https://doi.org/10.1007/JHEP06(2011)128}{\emph{JHEP} {\bfseries 06}
  (2011) 128} [\href{https://arxiv.org/abs/1106.0522}{{\ttfamily 1106.0522}}].

\bibitem{Alwall:2014hca}
J.~Alwall, R.~Frederix, S.~Frixione, V.~Hirschi, F.~Maltoni, O.~Mattelaer
  et~al., \emph{{The automated computation of tree-level and next-to-leading
  order differential cross sections, and their matching to parton shower
  simulations}}, \href{https://doi.org/10.1007/JHEP07(2014)079}{\emph{JHEP}
  {\bfseries 07} (2014) 079} [\href{https://arxiv.org/abs/1405.0301}{{\ttfamily
  1405.0301}}].

\bibitem{Aad:2015rba}
{\scshape ATLAS} collaboration, \emph{{Search for massive, long-lived particles
  using multitrack displaced vertices or displaced lepton pairs in pp
  collisions at $\sqrt{s}$ = 8 TeV with the ATLAS detector}},
  \href{https://doi.org/10.1103/PhysRevD.92.072004}{\emph{Phys. Rev. D}
  {\bfseries 92} (2015) 072004}
  [\href{https://arxiv.org/abs/1504.05162}{{\ttfamily 1504.05162}}].

\bibitem{Aaboud:2017iio}
{\scshape ATLAS} collaboration, \emph{{Search for long-lived, massive particles
  in events with displaced vertices and missing transverse momentum in
  $\sqrt{s}$ = 13 TeV $pp$ collisions with the ATLAS detector}},
  \href{https://doi.org/10.1103/PhysRevD.97.052012}{\emph{Phys. Rev. D}
  {\bfseries 97} (2018) 052012}
  [\href{https://arxiv.org/abs/1710.04901}{{\ttfamily 1710.04901}}].

\bibitem{Sjostrand:2014zea}
T.~Sj\"ostrand, S.~Ask, J.R.~Christiansen, R.~Corke, N.~Desai, P.~Ilten et~al.,
  \emph{{An introduction to PYTHIA 8.2}},
  \href{https://doi.org/10.1016/j.cpc.2015.01.024}{\emph{Comput. Phys. Commun.}
  {\bfseries 191} (2015) 159}
  [\href{https://arxiv.org/abs/1410.3012}{{\ttfamily 1410.3012}}].

\bibitem{Cacciari:2011ma}
M.~Cacciari, G.P.~Salam and G.~Soyez, \emph{{FastJet User Manual}},
  \href{https://doi.org/10.1140/epjc/s10052-012-1896-2}{\emph{Eur. Phys. J. C}
  {\bfseries 72} (2012) 1896}
  [\href{https://arxiv.org/abs/1111.6097}{{\ttfamily 1111.6097}}].

\bibitem{Cottin:2018kmq}
G.~Cottin, J.C.~Helo and M.~Hirsch, \emph{{Searches for light sterile neutrinos
  with multitrack displaced vertices}},
  \href{https://doi.org/10.1103/PhysRevD.97.055025}{\emph{Phys. Rev. D}
  {\bfseries 97} (2018) 055025}
  [\href{https://arxiv.org/abs/1801.02734}{{\ttfamily 1801.02734}}].

\bibitem{Degrande:2016aje}
C.~Degrande, O.~Mattelaer, R.~Ruiz and J.~Turner, \emph{{Fully-Automated
  Precision Predictions for Heavy Neutrino Production Mechanisms at Hadron
  Colliders}}, \href{https://doi.org/10.1103/PhysRevD.94.053002}{\emph{Phys.
  Rev. D} {\bfseries 94} (2016) 053002}
  [\href{https://arxiv.org/abs/1602.06957}{{\ttfamily 1602.06957}}].

\bibitem{ATLAS:2019kpx}
{\scshape ATLAS} collaboration, \emph{{Search for heavy neutral leptons in
  decays of $W$ bosons produced in 13 TeV $pp$ collisions using prompt and
  displaced signatures with the ATLAS detector}},
  \href{https://doi.org/10.1007/JHEP10(2019)265}{\emph{JHEP} {\bfseries 10}
  (2019) 265} [\href{https://arxiv.org/abs/1905.09787}{{\ttfamily
  1905.09787}}].

\bibitem{CMS:2018iaf}
{\scshape CMS} collaboration, \emph{{Search for heavy neutral leptons in events
  with three charged leptons in proton-proton collisions at $\sqrt{s} =$ 13
  TeV}}, \href{https://doi.org/10.1103/PhysRevLett.120.221801}{\emph{Phys. Rev.
  Lett.} {\bfseries 120} (2018) 221801}
  [\href{https://arxiv.org/abs/1802.02965}{{\ttfamily 1802.02965}}].

  \bibitem{CMS:2021lzm}
{\scshape CMS} collaboration, \emph{{Search for long-lived heavy neutral
  leptons with displaced vertices in pp collisions at
  $\sqrt{s}=13\,\mathrm{TeV}$ with the CMS detector}}, 
  \href{https://cds.cern.ch/record/2777047}{CMS-PAS-EXO-20-009} (2021).


\bibitem{Abreu:1996pa}
{\scshape DELPHI} collaboration, \emph{{Search for neutral heavy leptons
  produced in Z decays}}, \href{https://doi.org/10.1007/s002880050370}{\emph{Z.
  Phys. C} {\bfseries 74} (1997) 57}.

\bibitem{LHCb:2016inz}
{\scshape LHCb} collaboration, \emph{{Search for massive long-lived particles
  decaying semileptonically in the LHCb detector}},
  \href{https://doi.org/10.1140/epjc/s10052-017-4744-6}{\emph{Eur. Phys. J. C}
  {\bfseries 77} (2017) 224}
  [\href{https://arxiv.org/abs/1612.00945}{{\ttfamily 1612.00945}}].

\bibitem{Antusch:2017hhu}
S.~Antusch, E.~Cazzato and O.~Fischer, \emph{{Sterile neutrino searches via
  displaced vertices at LHCb}},
  \href{https://doi.org/10.1016/j.physletb.2017.09.057}{\emph{Phys. Lett. B}
  {\bfseries 774} (2017) 114}
  [\href{https://arxiv.org/abs/1706.05990}{{\ttfamily 1706.05990}}].

\bibitem{Drewes:2019fou}
M.~Drewes and J.~Hajer, \emph{{Heavy Neutrinos in displaced vertex searches at
  the LHC and HL-LHC}},
  \href{https://doi.org/10.1007/JHEP02(2020)070}{\emph{JHEP} {\bfseries 02}
  (2020) 070} [\href{https://arxiv.org/abs/1903.06100}{{\ttfamily
  1903.06100}}].

\bibitem{Davidson:1993qk}
S.~Davidson, D.C.~Bailey and B.A.~Campbell, \emph{{Model independent
  constraints on leptoquarks from rare processes}},
  \href{https://doi.org/10.1007/BF01552629}{\emph{Z. Phys. C} {\bfseries 61}
  (1994) 613} [\href{https://arxiv.org/abs/hep-ph/9309310}{{\ttfamily
  hep-ph/9309310}}].

\bibitem{PDG2020}
{\scshape Particle Data Group} collaboration, \emph{{Review of Particle
  Physics}}, {\emph{Prog. Theor. Exp. Phys.} {\bfseries 2020} (2020) 083C01}.

\bibitem{CMS:2018ncu}
{\scshape CMS} collaboration, \emph{{Search for pair production of
  first-generation scalar leptoquarks at $\sqrt{s} =$ 13 TeV}},
  \href{https://doi.org/10.1103/PhysRevD.99.052002}{\emph{Phys. Rev. D}
  {\bfseries 99} (2019) 052002}
  [\href{https://arxiv.org/abs/1811.01197}{{\ttfamily 1811.01197}}].

\bibitem{CMS:2018yiq}
{\scshape CMS} collaboration, \emph{{Search for dark matter in events with a
  leptoquark and missing transverse momentum in proton-proton collisions at 13
  TeV}}, \href{https://doi.org/10.1016/j.physletb.2019.05.046}{\emph{Phys.
  Lett. B} {\bfseries 795} (2019) 76}
  [\href{https://arxiv.org/abs/1811.10151}{{\ttfamily 1811.10151}}].

\bibitem{CMS:2019ybf}
{\scshape CMS} collaboration, \emph{{Searches for physics beyond the standard
  model with the $M_\mathrm{T2}$ variable in hadronic final states with and
  without disappearing tracks in proton-proton collisions at $\sqrt{s}=$ 13
  TeV}}, \href{https://doi.org/10.1140/epjc/s10052-019-7493-x}{\emph{Eur. Phys.
  J. C} {\bfseries 80} (2020) 3}
  [\href{https://arxiv.org/abs/1909.03460}{{\ttfamily 1909.03460}}].

\bibitem{ATLAS:2016wab}
{\scshape ATLAS} collaboration, \emph{{Search for scalar leptoquarks in pp
  collisions at $\sqrt{s}$ = 13 TeV with the ATLAS experiment}},
  \href{https://doi.org/10.1088/1367-2630/18/9/093016}{\emph{New J. Phys.}
  {\bfseries 18} (2016) 093016}
  [\href{https://arxiv.org/abs/1605.06035}{{\ttfamily 1605.06035}}].

\bibitem{ATLAS:2020dsk}
{\scshape ATLAS} collaboration, \emph{{Search for pairs of scalar leptoquarks
  decaying into quarks and electrons or muons in $ \sqrt{s} $ = 13 TeV $pp$
  collisions with the ATLAS detector}},
  \href{https://doi.org/10.1007/JHEP10(2020)112}{\emph{JHEP} {\bfseries 10}
  (2020) 112} [\href{https://arxiv.org/abs/2006.05872}{{\ttfamily
  2006.05872}}].

\bibitem{CMS:2014hka}
{\scshape CMS} collaboration, \emph{{Search for long-lived particles that decay
  into final states containing two electrons or two muons in proton-proton
  collisions at $\sqrt{s} =$ 8 TeV}},
  \href{https://doi.org/10.1103/PhysRevD.91.052012}{\emph{Phys. Rev. D}
  {\bfseries 91} (2015) 052012}
  [\href{https://arxiv.org/abs/1411.6977}{{\ttfamily 1411.6977}}].

\bibitem{CMS:2021kdm}
{\scshape CMS} collaboration, \emph{{Search for long-lived particles decaying
  to leptons with large impact parameter in proton-proton collisions at
  $\sqrt{s}$ = 13 TeV}},  \href{https://arxiv.org/abs/2110.04809}{{\ttfamily
  2110.04809}}.

\end{thebibliography}\endgroup
